\documentclass[aps,showpacs,superscriptaddress,nofootinbib,twocolumn,floatfix]{revtex4}
\usepackage[intlimits]{amsmath}
\usepackage{amssymb}
\usepackage{graphicx}
\usepackage[caption=false]{subfig}
\usepackage{siunitx}
\usepackage{booktabs}
\usepackage{mathrsfs,slashed}
\usepackage{nicefrac}
\usepackage[caption=false]{subfig}
\usepackage{float}

\begin{document}
\title{Effect of color superconductivity on the mass of hybrid neutron stars in an effective model with pQCD asymptotics }
\author{David~Blaschke}
\email{david.blaschke@uwr.edu.pl}
\affiliation{Institute of Theoretical Physics,
    University of Wroclaw,
    50-204 Wroclaw, Poland}
\author{Udita Shukla}
\email{udita.shukla28@gmail.com}
\affiliation{Institute of Theoretical Physics,
    University of Wroclaw,
    50-204 Wroclaw, Poland}
\author{Oleksii Ivanytskyi}
\email{oleksii.ivanytskyi@uwr.edu.pl}
\affiliation{Institute of Theoretical Physics,
    University of Wroclaw,
    50-204 Wroclaw, Poland}
\author{Simon~Liebing}
\email{science@liebing.cc}
\affiliation{Institute for Theoretical Physics, TU Bergakademie Freiberg, 09599 Freiberg, Germany}

\date{\today}
\begin{abstract}
The effective cold quark matter model by Alford, Braby, Paris and Reddy (ABPR) is used as a tool for discussing the 
effect of the size of the pairing gap in three-flavor (CFL) quark matter on the maximum mass of hybrid neutron stars (NSs). 
This equation of state (EOS) has three parameters which we suggest to determine by comparison with a nonlocal NJL model of quark matter in the nonperturbative domain. 
We show that due to the momentum dependence of the pairing which is induced by the nonlocality of the interaction, the effective gap parameter in the EOS model is well approximated by a constant value depending on the diquark coupling strength in the NJL model Lagrangian.
For the parameter $a_4=1-2\alpha_s/\pi$ a constant value below about \num{0.4} is needed to explain hybrid stars with 
${\rm M}_{\rm max} \gtrsim 2.0~{\rm M}_\odot$, which would translate to an effective constant $\alpha_s\sim 1$.
The matching point with a running coupling at the 1-loop $\beta$ function level is found to lie outside the range of chemical potentials accessible in NS interiors.
A dictionary is provided for translating the
free parameters 
of the nlNJL model to those of the ABPR model. Both models are shown to be equivalent in the nonperturbative domain but the latter one 
allows to quantify the transition to the asymptotic behaviour in accordance with perturbative QCD.
We provide constraints on parameter sets that fulfill the $2~{\rm M}_\odot$ mass constraint for hybrid NSs, 
as well as 
the low tidal deformability constraint from GW170817 by a softening of the EOS on the hybrid NS branch with an early onset of deconfinement at ${\rm M}_{\rm onset}<1.4~{\rm M}_\odot$.
We find that the effective constant pairing gap should be around 100 MeV but not exceed values of about 130 MeV because a further increase of the gap would entail a softening of the EOS and contradict the $2~{\rm M}_\odot$ mass constraint.
\end{abstract}

     
\maketitle

\section{Introduction}
Two limits of the cold dense matter equation of state (EOS) are precisely known: 
1) the state of nuclear matter at the nuclear saturation density $n_0=$\,\SI{0.15}{\per\femto\meter\tothe3} and below it; 
2) the cold quark matter EOS of perturbative QCD above about $40~n_0$. 
In between these limits the deconfinement phase transition has to take place. 
But the open question is whether its place could be in neutron star (NS) interiors.

By the end of the 1980's, the answer to this question by the authorities in the field was negative \cite{Bethe:1987sv} were it not for the possibility of exotic strange stars \cite{Witten:1984rs,Farhi:1984qu,Baym:1985tn,Alcock:1986hz,Alcock:1986xm,Haensel:1986qb}, made up of absolutely stable strange quark matter (SSQM) \cite{Baym:1988fa}.
At that time, works on stable quark matter cores in NSs like Ref. \cite{Blaschke:1989nn} not relying on the SSQM hypothesis but rather on assumptions for an interaction energy density functional were rather an exception. 
The nonrelativistic density functional of the confining string-flip model (SFM) \cite{Ropke:1986qs} that was used in \cite{Blaschke:1989nn} was recently generalized in a relativistic path integral formulation \cite{Kaltenborn:2017hus} which was successfully applied to study hybrid NSs, even forming a third family of compact stars \cite{Gerlach:1968zz,Benic:2014jia}. 

Among the density functionals for describing dense quark matter, those of the Nambu-Jona-Lasinio (NJL) type with relativistic current-current interactions obeying chiral symmetry but lacking confinement \cite{Klevansky:1992qe,Buballa:2003qv} have been widely used, also in considering the question of quark matter deconfinement in NS interiors.
These studies have shown that for a successful description of NS phenomenology with hybrid star sequences, two ingredients beyond the minimal NJL model interaction were essential: a vector meson channel for stiffening high-density quark matter, thus describing high-mass NSs and a scalar diquark interaction channel for lowering the onset of deconfinement \cite{Klahn:2006iw,Klahn:2013kga}.
For a recent review on the role of stiffness and color superconductivity in the description of the hadron-to-quark matter transition in NSs, see \cite{Baym:2017whm}.
To remedy the lack of confinement that limits the application of NJL-type models to the $T=$\,\num{0} region of the QCD phase diagram, the confining density functional approach has recently been developed to contain chiral symmetry and diquark interactions in the effective Lagrangian 
\cite{Ivanytskyi:2022oxv}. While this approach and its generalization to finite temperatures \cite{Ivanytskyi:2022wln} was still concerned with two quark flavors only, a special simplified version for three massless quark flavors in the color flavor locking (CFL) phase with a very early deconfinement transition triggered by light sexaquark condensation has been developed in \cite{Blaschke:2022knl}. 
The corresponding CFL phase with three degenerate light quark flavors may be called CFL-light (CFLL). 
It is worth mentioning, that an early transition to such a three-flavor CFL phase rather than to the two-flavor color superconducting quark matter is in line with the argument of high energy cost caused by imposing electric and color neutrality in the two-flavor case \cite{Alford:2002kj}.  
At the same time, electric neutrality of the CFL phase is provided automatically and does not cause an increase of its free energy \cite{Rajagopal:2000ff}.

Coming back to the present work, namely to join the nuclear matter phase with asymptotic perturbative QCD matter, we face the problem that the most advanced density functional approaches to cold quark matter (NJL and SFM) do not possess the pQCD limit, which may conveniently be characterized by approaching the conformal limit for the squared speed of sound, $c_s^2=1/3$, from below. 
The persistence of collective mean fields in the vector and diquark sector even at asymptotic densities makes NJL and SFM models violate the conformal limit. 
Since central densities in NSs reach only about
$5\,n_0$, one might argue that it shall not be a problem at all to construct a matching with the pQCD EOS at $40\,n_0$, which fulfills the basic constraints of causality ($c_s^2\le 1$) and thermodynamic stability.
But as it has been shown in \cite{Komoltsev:2021jzg}, there are EOS, e.g., in the CompOSE library of compact star EOS
\cite{CompOSECoreTeam:2022ddl} which do not allow such a matching unless it is introduced at sufficiently low densities and thus having an influence on the NS EOS.

Within the confining density functional approach, a procedure has been suggested that suggests a microscopic calculation  of the medium dependence of vector and diquark coupling constants using a massive gluon propagator ansatz, so that the conformal limit is restored \cite{Ivanytskyi:2022bjc}, see also \cite{Ivanytskyi:2022qnw}.

In the present work, we want to suggest another approach to 
define a quark matter EOS that unifies the requirement of a pQCD asymptotics at high densities with the nonperturbative features of confinement and color superconductivity in the region of the hadron-to-quark matter transition that likely takes place in the interior of light NSs and is advantageous for fulfilling modern multi-messenger constraints of NS phenomenology.  
The effective quark matter EOS suggested by Alford\,\emph{et al.} in \cite{Alford:2004pf} fulfills these conditions and at the same time has the advantage of simplicity that makes it suitable for extensive phenomenological studies.
We will use this form of EOS for color superconducting quark matter phases that was reused in several studies and in Ref.\,\cite{Zhang:2020jmb} given the form
\begin{align}
\label{eq:Pq1}
    P=\frac{\xi_4 a_4}{4\pi^2} \mu^4 + \frac{\xi_{2a}\Delta^2 -\xi_{2b} m_s^2}{\pi^2} \mu^2  
    - B_{\rm eff}~,
\end{align}
where the quark chemical potential $\mu$ is equivalently expressed through the baryon one $\mu_B=3 \mu$ and for the color-flavor-locking (CFL) phase holds that $\xi_4=$\,\num{3}, $\xi_{2a}=$\,\num{3} and $\xi_{2b}=$\,\nicefrac{3}{4}.
This model has the disadvantage that it uses four free parameters for which rather wide margins exist:
1) the coefficient $a_4=1-2\alpha_s/\pi$ that depends on the running fine structure constant of the strong interaction 
$\alpha_s$ in first order;
2) the diquark pairing gap $\Delta$;
3) the effective bag pressure $B_{\rm eff}$ and
4) the strange quark mass $m_s$.

It has, however, the advantage that it is very easy to use and allows to scan the space of opportunities for discussing color superconducting quark matter in NSs, bound to observational constraints for masses and radii.
For example, in Ref. \cite{Ozel:2010bz}, this model was employed in order to conclude immediately after the first Shapiro-delay based mass measurement on PSR J1614-2230 \cite{Demorest:2010bx} (which was revised in \cite{NANOGrav:2017wvv}) that a lower limit of \num{1.93}\,${\rm M}_\odot$ for maximum mass of NSs would entail that quark matter has to be strongly interacting ($a_4<0.63$) and color superconducting ($a_2 = m_s^2 - 4\Delta^2 < m_s^2$) when the onset of deconfinement is set to $1.5~n_0$ by an appropriate choice of $B_{\rm eff}$. 

In this work, we will consider the CFLL phase that allows to neglect the parameter $m_s$ and suggest to fix the remaining three parameters by fitting them to the diquark gap and the pressure of the nonlocal NJL model \cite{Schmidt:1994di,Contrera:2022tqh}.
In the latter, the scalar-pseudoscalar coupling, the light current quark mass and the range of the interaction are determined by low-energy vacuum QCD phenomenology, the pion mass and decay constant as well as the chiral condensate. The remaining unknown coupling constants in the vector meson and diquark interaction channels, $G_V$ and $G_D$, will be
mapped to the parameters of the ABPR model and can be constrained by the NS phenomenology.  
In this way, the present work will allow to link the parameters of an effective low-energy QCD Lagrangian to the effective ABPR quark matter model that is convenient to use in NS phenomenology and has the attractive feature of an asymptotic approach to the conformal limit which is in accordance with  pQCD.

\section{Color superconducting quark matter}
\subsection{The ABPR model}

\begin{figure}
    \centering
    \includegraphics[width=\columnwidth]{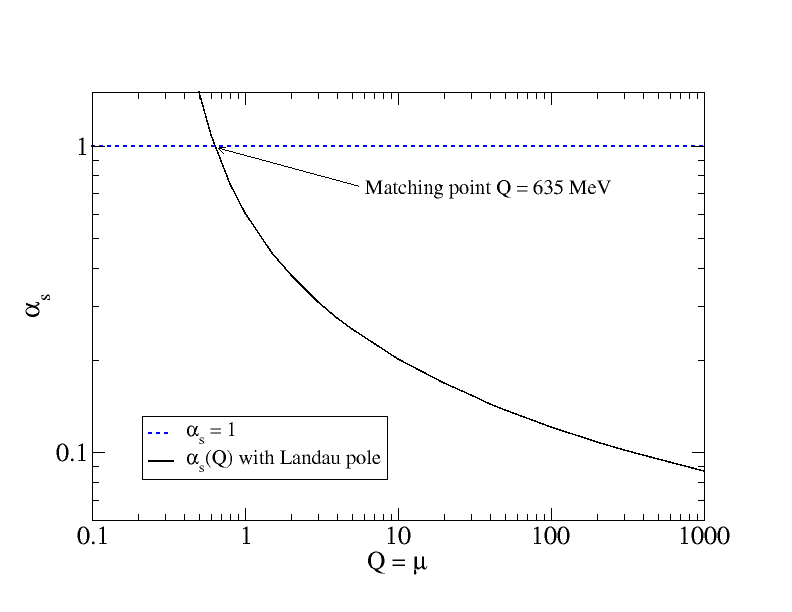}
    \caption{Running coupling $\alpha_s(Q=\mu)$ according to 1-loop $\beta$ function (black solid line) with a Landau pole at $Q=\Lambda=$\,\SI{315}{MeV} and a constant coupling at the level of the freezing value $\alpha_s(0)\sim 1$ (blue dotted line) which leads to a matching point at $Q=\mu=$\,\SI{635}{MeV}. 
    }
    \label{fig:alpha}
\end{figure}

In the following we will discuss a CFLL phase, where the strange quark has the same current mass as the up and down quarks, i.e. $m_u=m_d=m_s=m$. For practical purposes $m=0$ can be used as an excellent approximation. 
In this phase all quark species are equivalent and subject to the same paring gap $\Delta$. 
Then, their distribution functions become degenerate and consequently the partial densities are equal. 
This entails that the CFLL quark matter is neutral w.r.t. electric and color charges.
Therefore, no leptons will appear in CFLL matter. 
For the coefficient $a_4=1-2\alpha_s/\pi$ we will use two cases that are glued together at a matching point 
$\mu=\mu _B/3=$\,\SI{635}{MeV}, see Fig.~\ref{fig:alpha},
\begin{enumerate}
\item 
a constant value $\alpha_s=1$ corresponding to  $ a_4 =$\,\num{0.363}
in the non-perturbative domain for chemical potentials below the matching point, which are relevant for hybrid NSs, and
\item 
a running coupling according to the one-loop $\beta$ function of dense QCD, 
\begin{align}
\label{eq:alpha}
\alpha_s(\mu) = \frac{4\pi}{\beta_0 \ln (\mu^2/\Lambda^2)},
\end{align}
where $\beta_0=11-2N_f/3$ and $\Lambda=$\,\SI{315}{MeV}.
\end{enumerate}

\begin{figure}[htb]
    \includegraphics[width=0.9\linewidth]{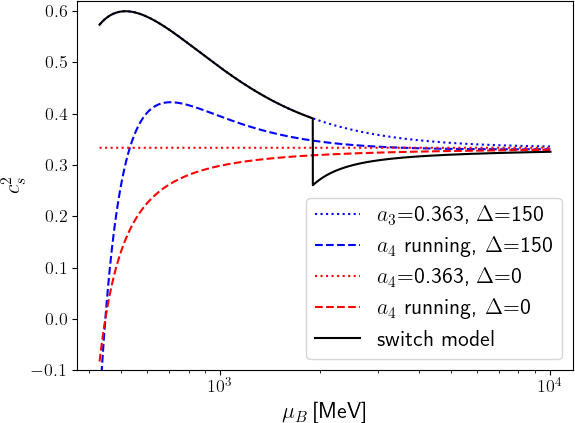}\\[2mm]
    \includegraphics[width=0.9\linewidth]{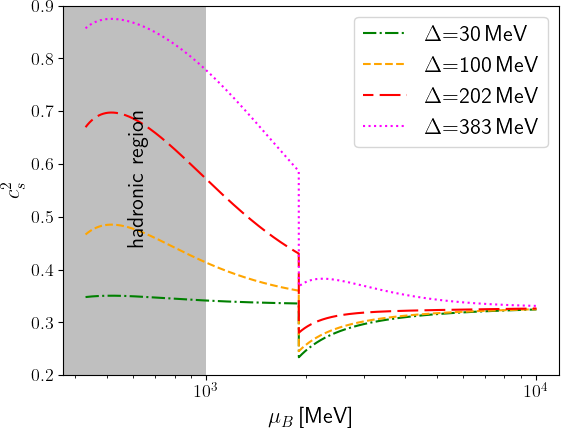}
    \caption{
    Upper panel: $c_s^2$ vs. $\mu_B$ for constant $a_4=$\,\num{0.363} and for running coupling, as well as for diquark gap $\Delta=$\,\SI{150}{MeV} and $\Delta=$\,\SI{0}.
    The solid black line corresponds to the instant switch model for $\alpha_s$ with a matching point at $\mu _B=1905$ MeV, see Fig.~\ref{fig:alpha}.
    Lower panel:
    $c_s^2$ vs. $\mu_B$ in the instant switch model for $\alpha_s$ for $a_4=$\,\num{0.363} and four cases of a constant diquark gap $\Delta=$\,\SIlist{30;100;202;383}{MeV}.}
    \label{fig:cs2}
\end{figure}

The resulting three-flavor, color superconducting quark matter EoS reads
\begin{align}
\label{eq:Pq2}
P=\frac{3 }{4\pi^2} a_4 \mu^4 + \frac{3}{\pi^2}\Delta^2 \mu^2 - B_{\rm eff}~.
\end{align}
For the quark number density follows
\begin{align}
\label{eq:n}
    n =& \frac{\partial P}{\partial \mu } \nonumber \\
    =& \frac{3 }{\pi^2} 
    \left(a_4 + \frac{33-2N_f}{12\pi^2}\alpha_s^2\right)\mu ^3+ 
    \frac{6}{\pi^2}\Delta^2\mu\,,
\end{align}
and the energy density is thus
\begin{align}
    \varepsilon =& \mu n - P
    \nonumber\\
    \begin{split}
         =& \frac{9}{4 \pi^2} 
\left(a_4 + \frac{33-2N_f}{9\pi^2}\alpha_s^2\right)\mu ^4
+ \frac{3}{\pi^2}\Delta^2 \mu^2+ B_{\rm eff}\,. 
    \end{split}
\end{align}
An interesting quantity is the squared sound speed which serves as a measure for the stiffness of the EOS.
It is obtained as
\begin{align}
\label{eq:cs2}
    c_s^2 &= \frac{d P}{d \varepsilon}
    = \frac{n}{\mu}\frac{d\mu}{dn}
    \nonumber\\   
    &= \frac{1+\zeta}{3+\zeta}\left[1-\frac{33-2N_f}{9\pi^2}\alpha_s^2\right] +\mathcal{O}(\alpha_s^3)\,,
\end{align}
where we introduced $\mu$-dependent function
\begin{align}
\label{eq:zeta}
    \zeta=\frac{18\Delta^2}{\mu _B ^2}\left[a_4 + \frac{33-2N_f}{12\pi^2}\alpha_s^2\right]^{-1}~.
\end{align}
We would like to discuss 
the two limiting cases of the matching model for a running $\alpha_s(\mu)$ shown in Fig. \ref{fig:alpha}.
\begin{enumerate}
    \item 
    For $a_4=1-2\alpha_s/\pi={\rm const.}$, the terms of higher order $\mathcal{O}(\alpha_s^2)$ in Eqs. (\ref{eq:n}) - (\ref{eq:zeta}), that arise from the $d\alpha_s/d\mu$ vanish and can thus
    be neglected.
    \item 
    For a running 
    $a_4(\mu)=1-2\alpha_s(\mu)/\pi$,
    the nonvanishing derivative w.r.t. the chemical potential 
    $\frac{d\alpha_s}{d\mu}=-\frac{33-2N_f}{6\pi}\frac{\alpha_s^2}{\mu}$ 
    results in the terms of $\mathcal{O}(\alpha_s^2)$, which then have to be taken into account.
\end{enumerate}
In the first case and for normal quark matter, when $\Delta=$\,\num{0}, the squared sound speed obeys the "conformal limit" value $c_s^2=$\,\nicefrac{1}{3}. 
Immediately after the deconfinement transition, when $\mu _B\approx \mu_c \approx$\,\SI{1150}{MeV} and for large diquark pairing gap, $\Delta\approx$\,\SI{150}{MeV}, the parameter $\zeta(\mu_c) \approx$\,\num{1} may be attained for $a_4=$\,\num{0.3} so that $c_s^2(\mu_c) =$\,\nicefrac{1}{2}. 
This value has been obtained as a typical result for several parametrizations of a nonlocal chiral quark model \cite{Antic:2021zbn,Contrera:2022tqh}.
The behavior of the squared sound speed in this case is shown in Fig.~\ref{fig:cs2}.

In the second case and for normal matter, $\zeta=0$, the squared sound speed approaches the "conformal limit" value $c_s^2=$\,\nicefrac{1}{3} from below for large $\mu$.
When in this case we consider a nonvanishing diquark gap, the squared sound speed has a profile similar to the one of the quarkyonic matter, rising to a peak above the conformal limit which is followed by a dip and asymptotically, for large $\mu$, approaching $c_s^2=$\,\nicefrac{1}{3} from below.

\subsection{Nonlocal NJL model for the CFLL phase}
\label{sec:nlNJL}

A simplified description of color superconducting three flavor quark matter assumes the same current mass $m$ of all quark flavors. 
Consequently, in such a model all quark flavors are  degenerate and have the same partial densities. This entails color and electric charge neutrality of this quark matter as well as a flavor independent quark chemical potential $\mu=\mu_B/3$.
In the case of the CFL phase, the microscopic states of paired quarks split into singlet and octet ones. 
The singlet states are characterized by the pairing gap of amplitude $2\Delta$, being twice the one of the octet states for which the pairing gap amplitude is $\Delta$. 
Below we label these states with the subscript index $j={\rm sing}$ for singlets and $j={\rm oct}$ for octets.
The assumption about the degeneration of the quark flavor states determines the CFLL phase.
In this work we model such phase of quark matter with a version of the nonlocal NJL model in the spirit of Ref. \cite{Contrera:2022tqh}.
Within this approach different interaction channels appear as correlations of quark currents, which are propagated through space via the three-momentum dependent form-factor $g_{\bf k}$.
In this work we adopt the Gaussian parameterization of $g_{\bf k}=\exp(-{\bf k}^2/\Lambda^2)$ where $\Lambda$ 
determines the finite range of the interaction in momentum space and replaces the constant cutoff parameter of the local NJL model.
We follow Ref. \cite{Contrera:2022tqh} and set $\Lambda=885.47$ MeV.
The chiral dynamics of the present model is represented by the melting of the mass gap amplitude $\sigma$ from some large vacuum value $\sigma_0$.
In what follows the subscript index ``$0$'' labels the quantities defined in the vacuum, i.e. at $\mu_B=0$. 
The mass gap amplitude enters the momentum dependent effective quark mass as $M_{\bf k}=m+\sigma_{\bf k}$ with $\sigma_{\bf k}\equiv\sigma g_{\bf k}$.
We adopt the current quark mass $m=2.29$ MeV from Ref. \cite{Contrera:2022tqh}.
Effects of the vector repulsion are controlled by the zeroth component of the  vector meson field (see Refs. \cite{Klahn:2006iw,Contrera:2022tqh} for details), 
i.e. by $\omega$. 
Similar to Ref. \cite{Contrera:2022tqh}, we assume the corresponding vector interaction channel to be local, which formally corresponds to the formfactor $g_{\bf k}^V=1$. 
This simplification allows us to avoid serious technical complications in evaluating the Matsubara sums that would be caused by the appearance of the Matsubara index in the expression for the effective chemical potential of quarks.

With the above notations, the  zero temperature thermodynamic potential of the CFLL phase can be written as
\begin{align}\label{eq:Omega}
\Omega=&\frac{\sigma^2}{4G_S}-\frac{\omega^2}{4G_V}
+\frac{\Delta^2}{4G_D}+\Omega_q.
\end{align}
Here $G_S$, $G_V$ and $G_D$ stand for couplings in the scalar-pseudoscalar, vector and diquark channels, respectively, and $\Omega_q$ is the quark contribution to the thermodynamic potential. 
Two of the above couplings are parameterized by the dimensionless quantities $\eta_V\equiv G_V/G_S$ and $\eta_D\equiv G_D/G_S$.
Parameterizations of the CFLL EOS considered below are labeled with pairs of numbers $(\eta_V,\eta_D)$. For example, $(1.3,0.6)$ corresponds to an EOS obtained for $G_V=1.3\, G_S$ and $G_D=0.6\, G_S$.
The quark term in $\Omega$ is
\begin{align}\label{eq:Omegaq}
\Omega_q=-\sum_{j,a=\pm}d_j\int\frac{d{\bf k}}{(2\pi)^3}
\left[\frac{\epsilon_{j{\bf k}}^a}{2}
-\epsilon_{j{\bf k}}^a f_{j{\bf k}}^a\right].
\end{align}
The summation in this expression is performed over singlet and octet quark ($a=+$) and antiquark ($a=-$) states.
The degeneracy factors of the singlet and octet states are expressed through the spin-flavor-color degeneracy one $d=2\times3\times3$ as $d_{\rm sing}=d/9$ and $d_{\rm oct}=8d/9$.
The corresponding single particle distribution function 
$f_{j{\bf k}}^a=\theta(-\epsilon_{j{\bf k}}^a)$ is given in terms of the single particle energy $\epsilon_{j{\bf k}}^a$ shifted by the effective chemical potential $\mu^\pm=\pm\mu\mp\omega$.
For definiteness we consider positive baryonic chemical potentials leading to $\mu^+>0$ and $\mu^-<0$. 
Thus
\begin{align}
\label{eq:dispersion}
\epsilon_{j{\bf k}}^a={\rm sgn}(\epsilon_{\bf k}-\mu^a)\sqrt{(\epsilon_{\bf k}-\mu^a)^2+\Delta_{j{\bf k}}^2},
\end{align}
where $\epsilon_{\bf k}=\sqrt{{\bf k}^2+M_{\bf k}^2}$ is the single particle energy of unpaired quarks, while $\Delta_{j{\bf k}}=\zeta_j\Delta g_{\bf k}$ with $\zeta_{\rm sing}=2$ and $\zeta_{\rm oct}=1$ is introduced in order to unify the notations. 
At $\epsilon_{\bf k}=\mu^+$, which defines the Fermi momentum $k_F$, $\epsilon_{j{\bf k}}^+$ experiences a discontinuous jump of the amplitude $2\Delta_{j{\bf k}}|_{|{\bf k}|=k_F}$, which is twice the gap of the energy spectrum of the corresponding quark state.
In order to quantify it we introduce the effective pairing gap
\begin{align}
\label{eq:gapeff}
\Delta_{\rm eff}\equiv\Delta g_{\bf k}|_{|{\bf k}|=k_F}.
\end{align}

The first term in the square brackets in Eq. (\ref{eq:Omegaq}) corresponds to the divergent zero point contribution to the thermodynamic potential.
It can be regularized by subtracting the constant vacuum value of $\Omega$.
This leads to the regularized thermodynamic potential
\begin{align}
\label{eq:Omegareg}
\Omega_{\rm reg}=\Omega-\Omega_0.
\end{align}

The physical values of the mass gap and pairing gap amplitudes as well as the zeroth component of the vector field can be found by minimizing the regularized thermodynamic potential with respect to $\sigma$, $\omega$ and $\Delta$.
This yields
\begin{align}
\label{eq:massgap}
\sigma=&2G_S\sum_{j,a=\pm}d_j\int\frac{d{\bf k}}{(2\pi)^3}
\left(\frac{1}{2}-f_{j{\bf k}}^a\right)
\frac{\epsilon_{\bf k}-\mu^a}{\epsilon_{j{\bf k}}^a}
\frac{M_{\bf k} g_{\bf k}}{\epsilon_{\bf k}},\hspace*{.5cm}\\
\label{eq:omega}
\omega=&2G_V\sum_{j,a=\pm}d_j\int\frac{d{\bf k}}{(2\pi)^3}
\left(\frac{1}{2}-f_{j{\bf k}}^a\right)a
\frac{\epsilon_{\bf k}-\mu^a}{\epsilon_{j{\bf k}}^a},\\
\label{eq:gap}
\Delta=&2G_D \sum_{j,a=\pm}d_j\int\frac{d{\bf k}}{(2\pi)^3}
\left(\frac{1}{2}-f_{j{\bf k}}^a\right)
\frac{\Delta}{\epsilon_{j{\bf k}}^a}
\zeta_j^2g_{\bf k}^2.
\end{align}
The equations for the amplitudes of the mass and pairing gaps include the zero point terms, which are regular due to the presence of the form-factor under the corresponding momentum integrals.
It is also worth mentioning that contrary to the case of local current interaction (see, e.g., Refs. \cite{Blaschke:2005uj,Klahn:2006iw}), Eqs. (\ref{eq:massgap}), (\ref{eq:omega}) do not include the terms with the Dirac delta-function $\delta(\epsilon_{\bf k}-\mu^a)$ arising from differentiating ${\rm sgn}(\epsilon_{\bf k}-\mu^a)$ in the dispersion relation (\ref{eq:dispersion}).
This is due to $f_{j{\bf k}}^a=1/2$ at vanishing $\epsilon_{j{\bf k}}^a=0$ providing zero value of the factor $1/2-f_{j{\bf k}}^a$ under the momentum integrals.
Having Eqs. (\ref{eq:massgap}) - (\ref{eq:gap}) solved we can construct EOS of the CFLL phase by defining its pressure $P=-\Omega_{\rm reg}-\Delta B$, baryon density $n_B=dP/d\mu_B=\omega/6G_V$, energy density $\varepsilon=\mu_Bn_B-P$ and squared speed of sound $c_S^2=dP/d\varepsilon$.
The contribution $-\Delta B$ to $P$ is a phenomenological constant pressure shift that could be motivated by a medium dependence in the nonperturbative gluon background (confinement) that is not captured by the nlNJL model for the quark dynamics. Such a constant has been introduced, e.g., in Refs. \cite{Pagliara:2007ph,Blaschke:2010vj,Bonanno:2011ch}
in order to regulate the onset density of quark deconfinement.

Before going further we would like to consider the question about the instability of the vacuum with respect to formation of the color superconducting state. 
This happens if the diquark coupling exceeds some critical value $G_D^*$.
At $G_D=G_D^*$ the second order phase transition to the CFLL phase occurs in the vacuum. 
In other words, $\partial^2\Omega/\partial\Delta^2=0$ at $\Delta=0$ in the vacuum ($f_{j{\bf k}}^a=0$).
This allows us to find the critical value of the diquark coupling as
\begin{align}
\label{GDcrit}
G_D^*=\left[\frac{8d}{3}\int\frac{d{\bf k}}{(2\pi)^3}\frac{g_{\bf k}^2}{\epsilon_{\bf k}}\right]^{-1}
\end{align}
parameterized via $\eta_D^*\equiv G_D^*/G_S$.
This expression includes the vacuum value of the mass gap amplitude $\sigma_0$ defined by Eq. (\ref{eq:massgap}) under the conditions $\Delta=0$ and $\mu^a=0$.
Adjusting the scalar coupling $G_S=3.307~{\rm GeV}^{-2}$ so that $\sigma_0=330$ MeV, we obtain $\eta_D^*=0.761$.
In order to provide vacuum stability we limit our analysis to the values of the diquark coupling respecting the requirement $\eta_D\le0.75$.

\begin{figure}
\centering
\subfloat{\hspace{-2em}
\includegraphics[width=0.9\columnwidth]{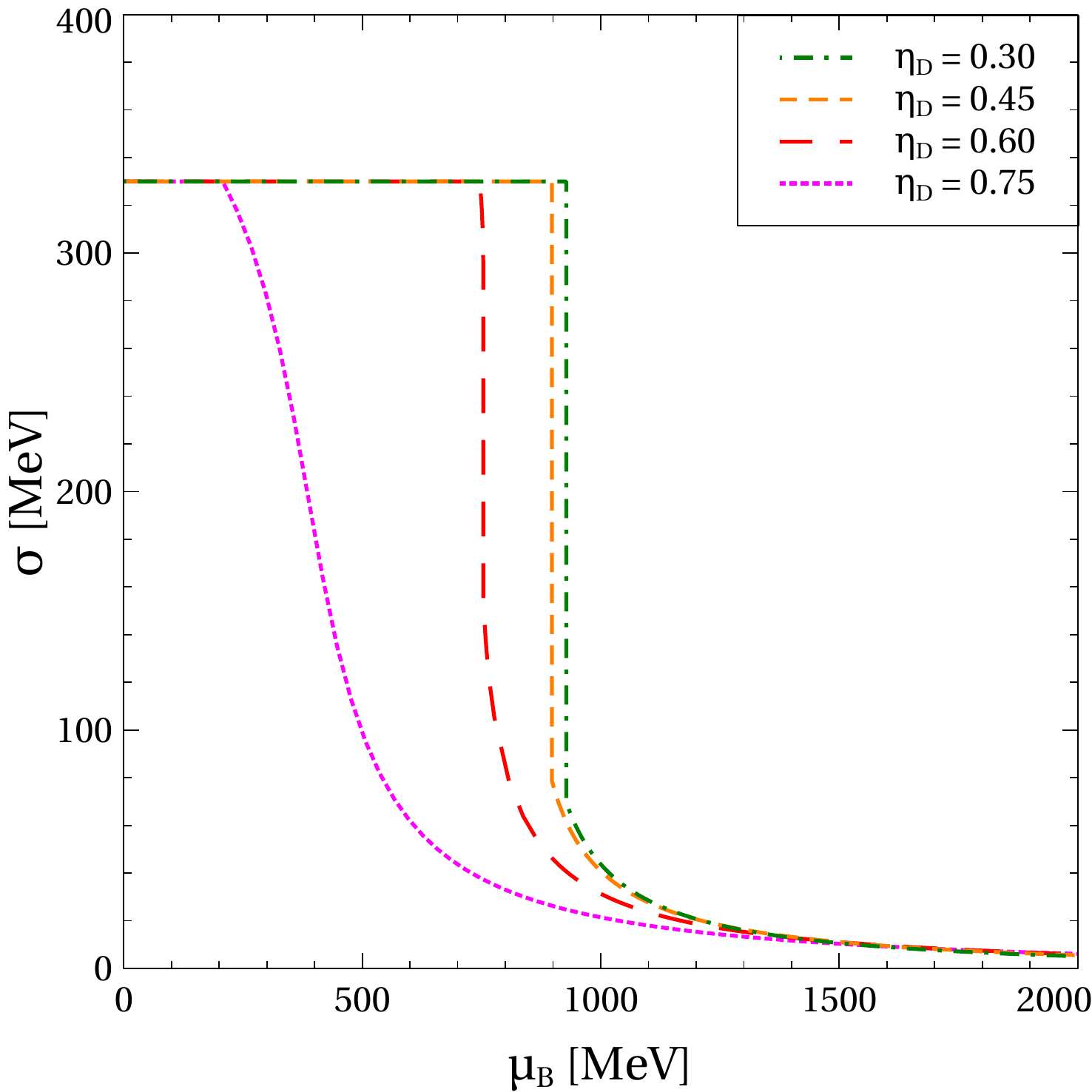}}\\
\subfloat{\hspace{-2em}
\includegraphics[width=0.9\columnwidth]{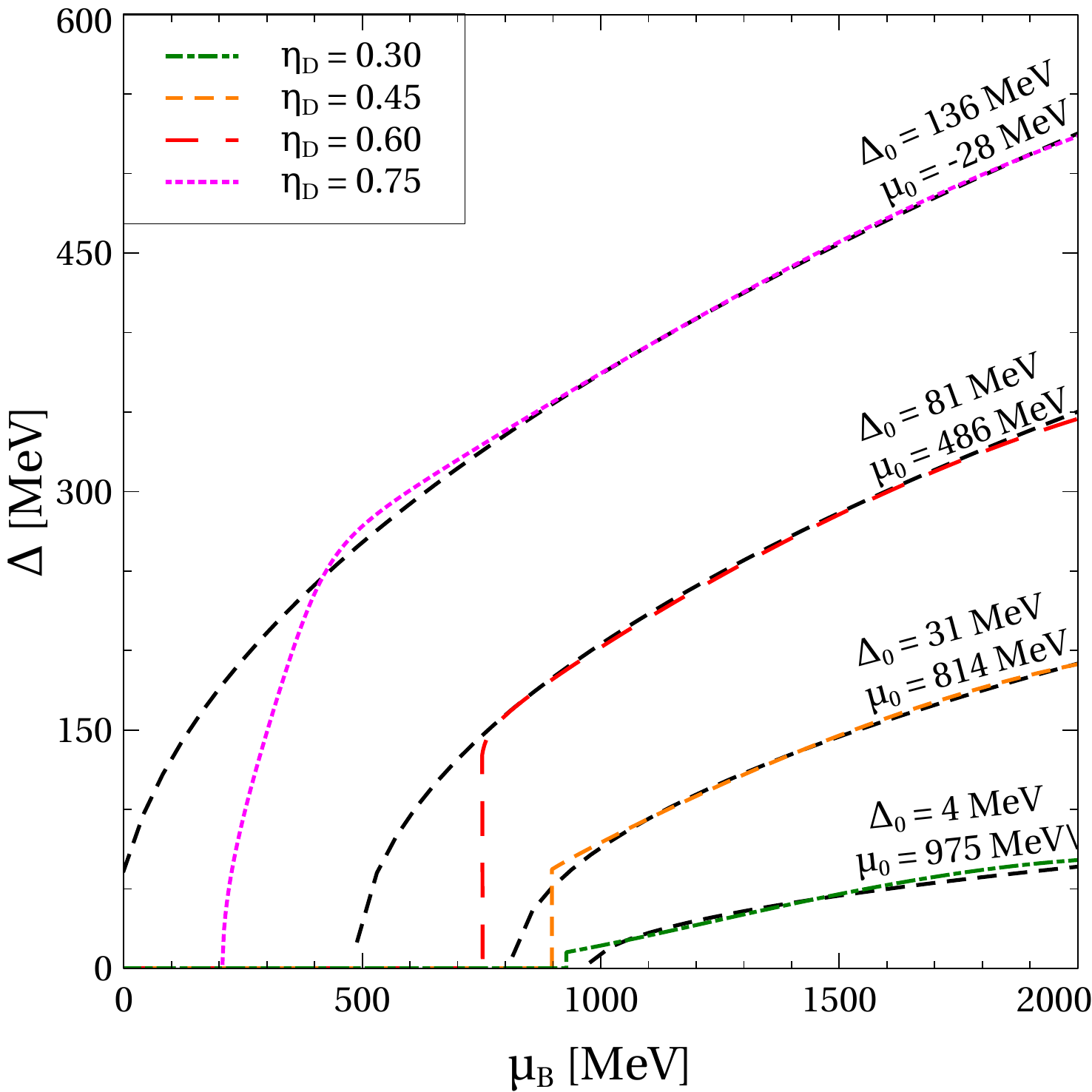}}
\caption{Amplitude of the mass gap $\sigma$ (upper panel) and diquark gap $\Delta$ (lower panel) as functions of baryonic chemical potential $\mu_B$ calculated for vanishing vector coupling $\eta_V=0$ and diquark couplings $\eta_D$ given in the legends and $\Delta B=0$. The black dashed curves on the lower panel represent the results of fitting the pairing gap amplitude by $\Delta=\sqrt{\Delta_0(\mu_B-\mu_0)}$ as discussed in the text.}
\label{fig:gap}
\end{figure}

The amplitudes of the pairing gap $\Delta$ and the mass gap $\sigma$ as functions of the baryonic chemical potential are shown in Fig. \ref{fig:gap} for different choices of the diquark coupling $\eta_D$.
At small but finite value of the baryon chemical potential $\sigma$ has its vacuum value and $\Delta$ vanishes. 
This corresponds to the chirally broken normal phase of heavy unpaired quarks.
At a certain value of $\mu_B$ the mass gap amplitude starts to melt, while the pairing gap amplitude starts to grow.
At small $\eta_D\ll1$ this happens discontinuously, with a jump in the chiral ($\sigma$) and color superconductivity ($\Delta$) order parameters, signalling a first order phase transition to the CFLL quark matter. 
At large $\eta_D\simeq1$, however, this transition is of the second order for the diquark condensate, with the onset starting from $\Delta=0$ followed by a continuous increase that goes over to a square root behaviour 

\begin{equation}
\label{eq:Delta}
    \Delta=\sqrt{\Delta_0(\mu_B-\mu_0)}
\end{equation}
with $\Delta_0$ and $\mu_0$ being constant parameters.
For a given value of the diquark coupling these parameters are defined by fitting the pairing gap amplitude within the range of the baryon chemical potentials from $\mu_1=$\,\SI{1}{GeV} to $\mu_2=$\,\SI{2}{GeV}, which covers the values typical for quark matter in NSs.
On the lower panel of Fig. \ref{fig:gap}, we demonstrate the quality of the fit (\ref{eq:Delta}) and provide the parameters $\Delta_0$ and $\mu_0$ for four cases of diquark couplings $\eta_D$. 

Above the color superconductivity onset this square root dependence perfectly describes behavior of $\Delta$.
It is interesting to note that this scaling law for the diquark condensate was found in a recent study of lattice simulations of two-color QCD \cite{Iida:2019rah}.

\begin{figure}
\centering
\includegraphics[width=0.9\columnwidth]{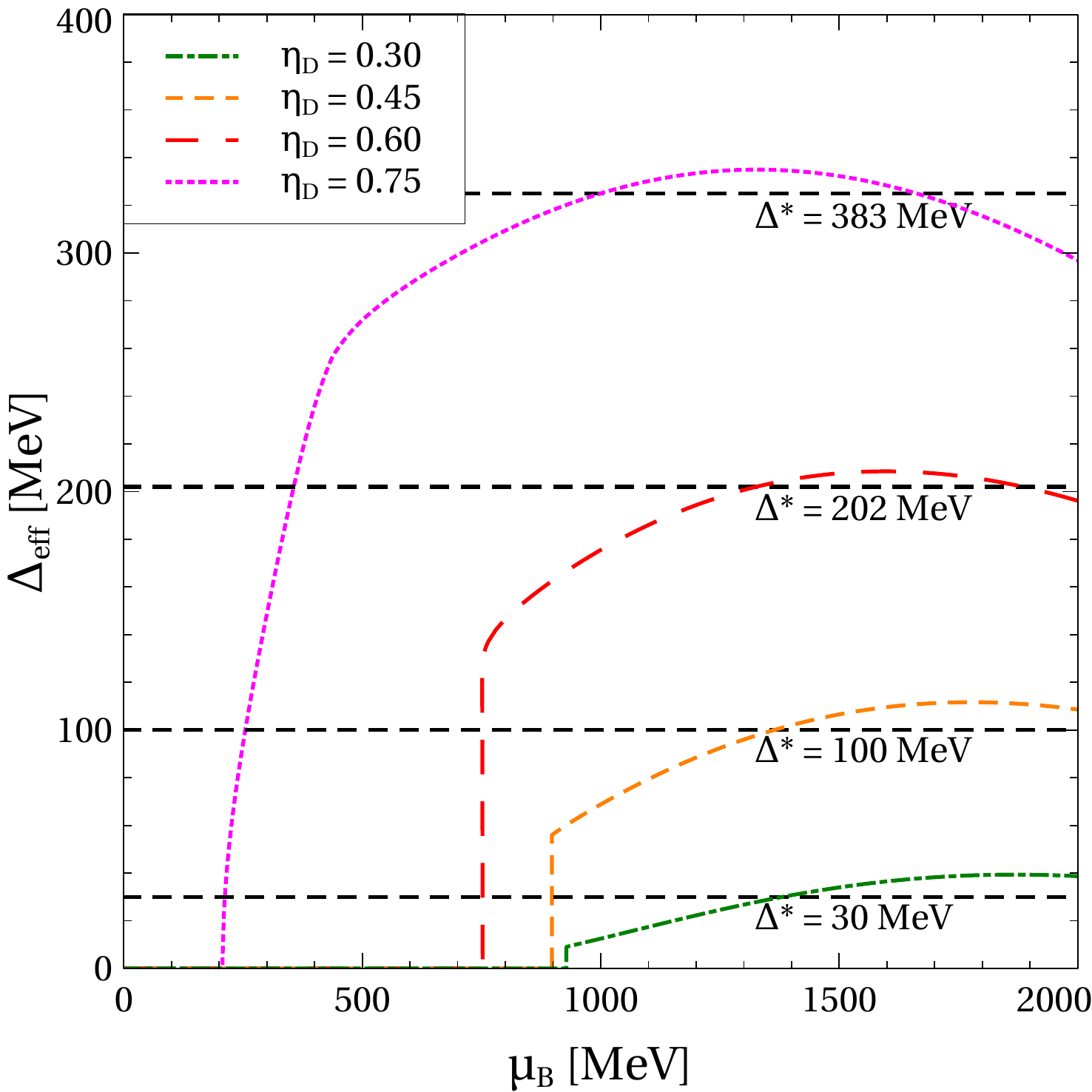}
\caption{Effective pairing gap $\Delta_{\rm eff}$ as a function of baryonic chemical potential $\mu_B$ calculated for vanishing vector coupling $\eta_V=0$ and diquark couplings $\eta_D$ given in the legend.
The black dashed lines represent the average values of the effective pairing gap $\Delta^*$ discussed in the text.}
\label{fig:gapeff}
\end{figure}

While $\Delta$ grows with $\mu_B$, the momentum form-factor $g_{\bf k}$ defined at $|{\bf k}|=k_F$ exhibits the opposite behavior. 
As a result the effective paining gap $\Delta_{\rm eff}$ significantly flattens.
Its behavior is shown in Fig. \ref{fig:gapeff}.
Within the range of baryon chemical potentials from $\mu_1$ to $\mu_2$, the effective pairing gap can be approximated by its average value
\begin{align}
\label{eq:gapaverage}
\Delta^*=\frac{1}{\mu_2-\mu_1}\int_{\mu_1}^{\mu_2}d\mu_B\Delta_{\rm eff}(\mu_B).
\end{align}
It is worth mentioning that this average value is almost insensitive to the vector coupling, which does not impact $\Delta_{\rm eff}$ but simply renormalizes $\mu_B$. 
Larger diquark couplings lead to stronger quark pairing and, consequently, to larger average values of the effective pairing gap of the CFLL matter.

\begin{figure}[t]
    \centering
    \vspace{-9mm}
    \includegraphics[width=0.55\textwidth]{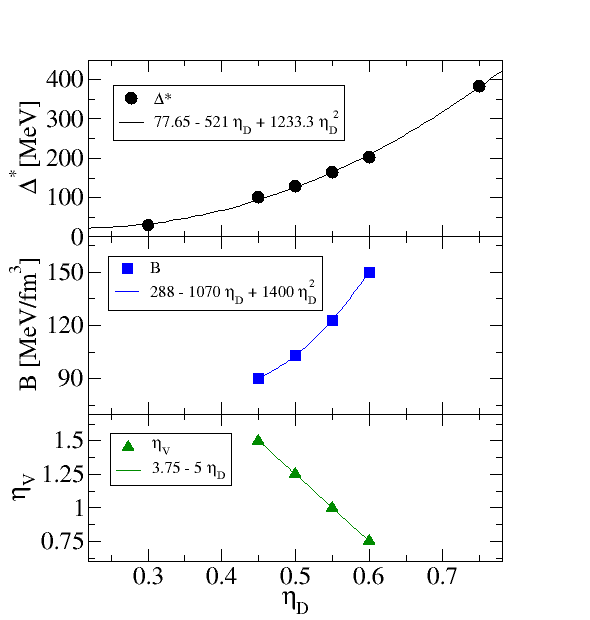}
    \caption{
    Dependence of the effective constant diquark gap $\Delta^*$ (top panel) and 
    the bag pressure $B$ (middle panel) of the ABPR model on the diquark coupling $\eta_D$ in the nlNJL model Lagrangian. The bottom panel shows the vector meson coupling $\eta_V$ required for the matching between nlNJL and ABPR model in the low-energy domain of neutron star chemical potentials.
    }
    \label{fig:Delta}
\end{figure}

For the functional dependence of $\Delta^*$ on $\eta_D$, we found a quadratic fit, see the upper panel of
Fig.~\ref{fig:Delta}, 
\begin{align}
\label{eq:Delta*}
    \Delta^*[{\rm MeV}] =77.65 - 521\, \eta_D + 1233.3\, \eta_D^2~.
\end{align}

\begin{figure}
\centering
\subfloat{\hspace{-2.5em}
    \includegraphics[width=0.85\columnwidth]{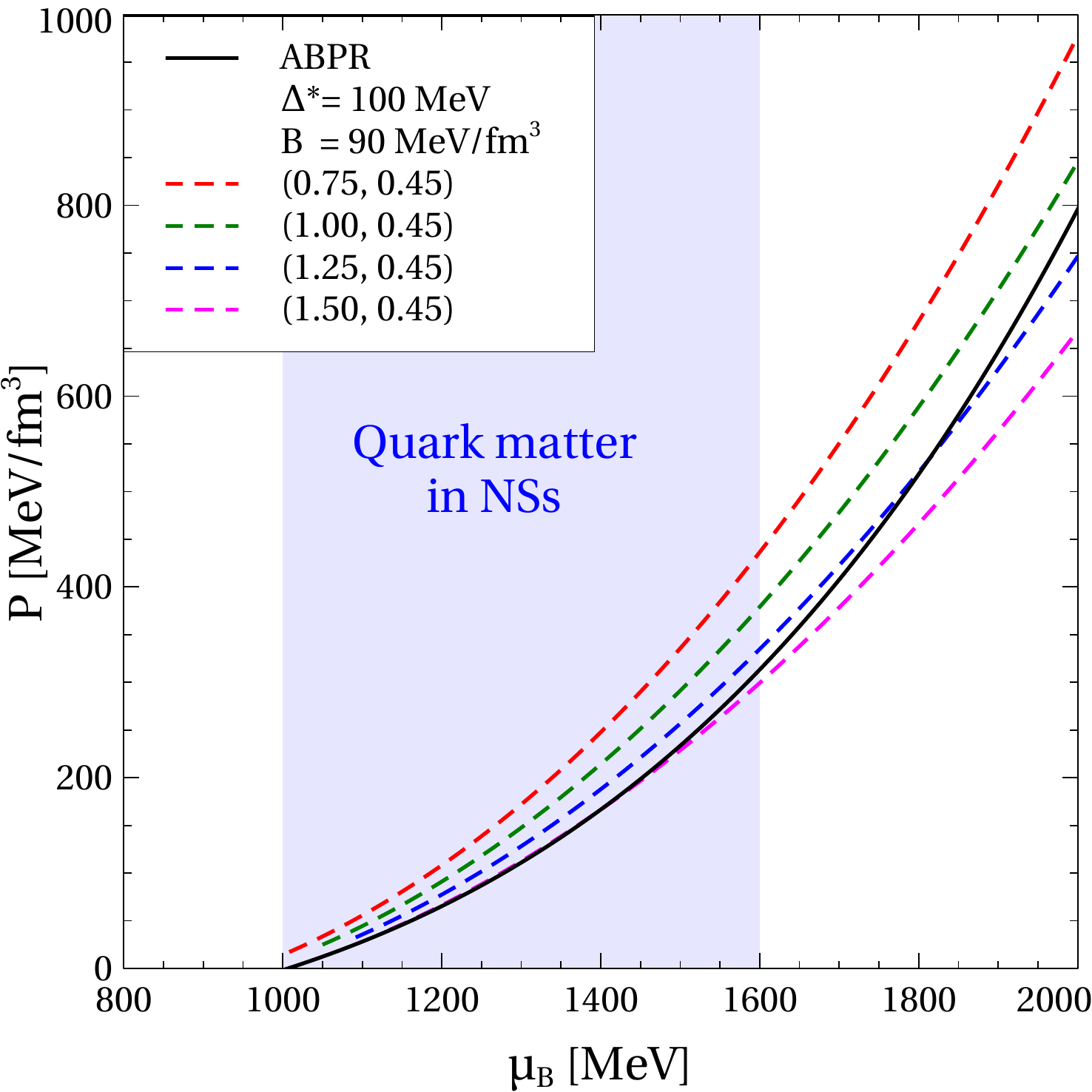} }\\\vspace{-1em}
 \subfloat{\hspace{-2.5em}
    \includegraphics[width=0.85\columnwidth]{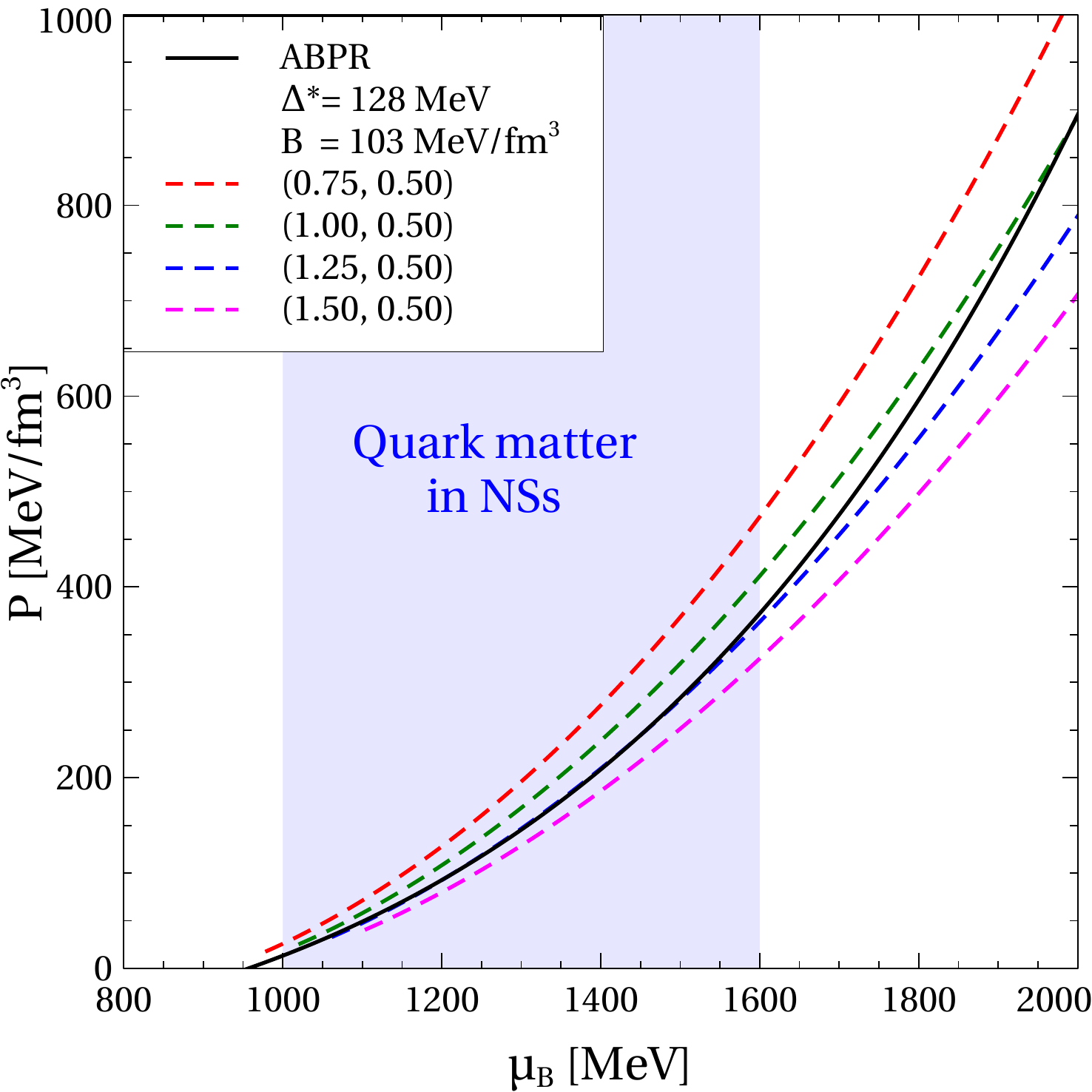} }\\\vspace{-1em}   
 \subfloat{\hspace{-2.5em}
     \includegraphics[width=0.85\columnwidth]{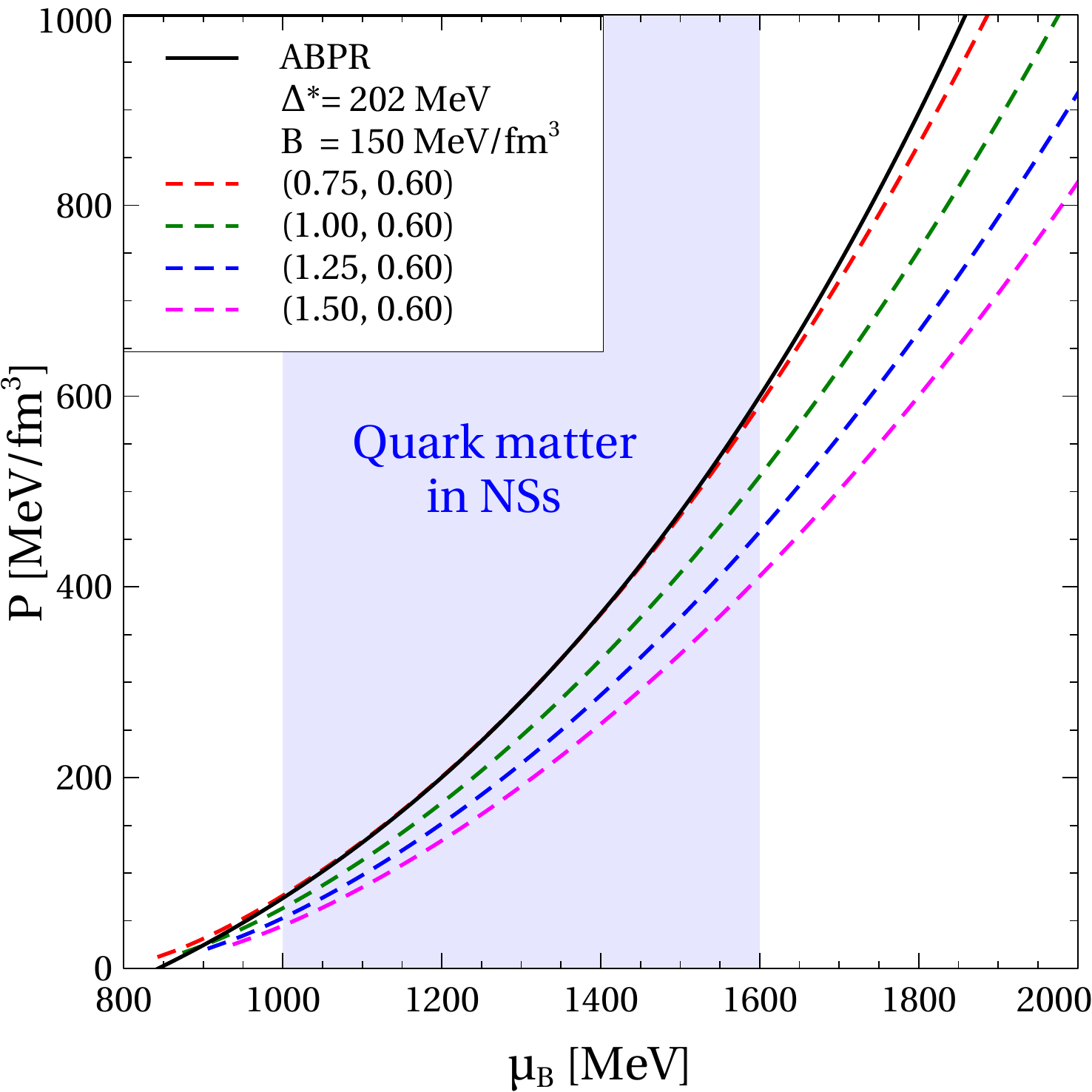}  }
\caption{Pressure of the CFLL quark matter $p$ as function of 
baryonic chemical potential $\mu_B$ calculated for several values of $\eta_V$ indicated in the legends and $\eta_D=$\,\num{0.45} (top panel), $\eta_D=$\,\num{0.5} (middle panel) and $\eta_D=$\,\num{0.60} (bottom panel).
The results are compared to the pressure of the ABPR model with the pairing gap $\Delta^*$
and bag pressure $B_{\rm eff}$, which coincides with $B$ for $\Delta B=0$
shown here.
The shaded area represents the approximate range of $\mu_B$ typical for quark matter in NSs.}
\label{fig:pessure}
\end{figure}

On the other hand, $\eta_V$ regulates the     stiffness of the CFLL EOS being in one-to-one correspondence with the slope of pressure defined as a function of $\mu_B$.
As is seen from Fig. \ref{fig:pessure}, within the range of the baryonic chemical potentials typical for quark matter in the cores of NS the vector coupling can be adjusted so that the pressure slopes of the CFLL quark matter and ABPR model coincide.
We would like to stress, the stronger is the diquark pairing the weaker should be the vector repulsion providing the same slopes of the CFLL and ABPR pressures.
Coincidence of the absolute values of pressures of these two models requires an adjustment of the bag pressure $B$ of the 
ABPR model, which grows with the diquark coupling, 
see the lower panel of Fig. \ref{fig:Delta}.
Also for this behaviour a quadratic fit is found
\begin{align}
\label{eq:B}
    B[{\rm MeV/fm}^3] = 288 - 1070\, \eta_D + 1400\, \eta_D^2~.
\end{align}
To summarize, we note that for a fixed choice 
of the QCD structure constant $\alpha_s$, 
the parameterization of the ABPR model that would provide a matching with the 
nlNJL model describing the CFLL quark matter
is determined by the choice of the diquark coupling $\eta_D$. 
There is a one-to-one correspondence of this parameter to the constant pairing gap $\Delta^*$ expressed by Eq. (\ref{eq:Delta*}).
With a proper choice of $\eta_V$ that reproduces the curvature of the ABPR pressure in the NS density range, the constant shift $B$ follows unambiguously. 
There is a linear relationship between $\eta_V$ and $\eta_D$
\begin{align}
    \eta_V=3.75 - 5.0\, \eta_D~.
\end{align}
The complete dictionary can be read-off
Fig. \ref{fig:Delta}.
This motivates us to establish a connection between relatively simple phenomenological ABPR EOS and microscopic nlNJL approach to the CFLL quark matter. 


\subsection{Parameter matching between ABPR and nlNJL models}
\label{ssec:matching}

The ABPR model \cite{Alford:2004pf} is built on top of the bag model with pQCD corrections by taking into account the effects of quark pairing in a perturbative manner.
The corresponding pressure reads
\begin{align}
\label{eq:ABPR}
P_{\rm ABPR} = P_{\rm free}+P_{\rm pert} + P_{\rm pair} - B_{\rm eff}
\end{align}
\begin{table*}[!th]
    \centering
    \begin{tabular}{c||c|c||c|c|c||c|c|c|c|c|c|c}
        set & $\eta_D$ & $\eta_V$ & $\Delta^*$ & $B$ & $\Delta B$ & $\rm M_{\rm onset}$ & $\rm M_{\rm max}$ & $\mu_B^{\rm max}$ & $\varepsilon^{\rm max}$ & $\rm R_{2.0}$ & $\rm R_{1.4}$ & $\Lambda_{1.4}$ \\
        &&&[MeV]&[MeV/fm$^3$]&[MeV/fm$^3$]& [$\rm M_\odot$]& [$\rm M_\odot$]&
        [MeV]& MeV/fm${}^3$&[km] & [km]\\
        \hline
        \hline
I a &0.45&1.5&100&90& -7  & 1.37 & 2.19 & 1597 & 1020 & 13.05 & 13.17 & 678\\
I b &0.45&1.5&100&90& -8  & 0.84 & 2.20 & 1578 & 1016 & 12.99 & 13.04 & 623\\
I c &0.45&1.5&100&90& -11 & 0.24 & 2.23 & 1570 & 996  & 12.85 & 12.57 & 565\\
\hline
II a &0.5 &1.25&128&103& 15 & 1.37 & 2.03 & 1617 & 1217 & 11.48 & 13.13 & 660 \\
II b &0.5 &1.25&128&103&  9 & 0.84 & 2.05 & 1624 & 1226 & 11.77 & 12.32 & 402 \\
II c &0.5 &1.25&128&103&  2 & 0.24 & 2.10 & 1604 & 1172 & 11.74 & 11.65 & 349 \\
\hline
III a &0.6&0.75&202&150& 102& 1.37 & 1.64 & 1740 & 2014 & -- & 11.10& 148\\
III b &0.6&0.75&202&150& 77 & 0.84 & 1.69 & 1741 & 1994 & -- & 9.96 & 81 \\
III c &0.6&0.75&202&150& 57 & 0.24 & 1.78 & 1695 & 1822 & -- & 9.43 & 88 \\
\hline
\hline
   \end{tabular}
    \caption{Parameter matching between nlNJL model (columns 2 and 3) and the ABPR model (columns 3 and 4) for the chosen case $a_4=$\,\num{0.363}. The parameter $\Delta B$ is added to both EOS to fix the onset of deconfinement.}
    \label{tab:param}
\end{table*}
The effective bag pressure 
$B_{\rm eff}=B+\Delta B$, which is present in Eqs. (\ref{eq:Pq1}) and (\ref{eq:Pq2})
absorbs the two constant parameters $B$ and $\Delta B$.
It is worth mentioning that $B$ provides consistency of the ABPR and nlNJL EOSs, 
while $\Delta B$ is introduced 
to adjust the onset density of quark deconfinement.

The negative of the regularized thermodynamic potential (\ref{eq:Omegareg}) with $m=\sigma=\omega=\Delta=0$ yields the pressure of free massless quarks with the single particle energies shifted by the chemical potential $\mathfrak{e}_{\bf k}=|{\bf k}|-\mu$ and the distribution function being $\mathfrak{f}_{\bf k}=\theta(-\mathfrak{e}_{\bf k})$.
Thus
\begin{align}
\label{eq:ABPR1}
P_{\rm free}=-d\int \frac{d{\bf k}}{(2\pi)^3}\mathfrak{e}_{\bf k}\,
\mathfrak{f}_{\bf k}=\frac{3}{4\pi^2} \mu^4.
\end{align}
The order $\mathcal{O}(\alpha_s)$ perturbative correction $P_{\rm pert}$ can be obtained as a two-loop exchange energy of massless quarks \cite{Kapusta:1989tk}.
For $N_f=3$ flavors and $N_c=3$ colors it is
\begin{align}
P_{\rm pert}=-\frac{N_f(N_c^2-1)g^2}{4}
\left[\int\frac{d{\bf k}}{(2\pi)^3}\frac{\mathfrak{f}_{\bf k}}{\mathfrak{e}_{\bf k}}\right]^2
\label{eq:ABPR2}
=-\frac{3\, \alpha_s}{2 \pi^3} \mu^4.\,
\end{align}
Here the QCD structure constant $\alpha_s=g^2/4\pi$ is expressed through the 
QCD coupling $g$. 
The effects of quark pairing can be taken into account by introducing $P_{\rm pair}$ being the negative of $\Omega_{\rm reg}$ with $m=\sigma=\omega=0$ and subtracted $P_{\rm free}$, i.e.
\begin{align}\label{eq:PCFLL1}
P_{\rm pair}=
\sum_jd_j\int \frac{d{\bf k}}{(2\pi)^3}
\sqrt{\mathfrak{e}_{\bf k}^2+\Delta_{j{\bf k}}^2}\,
\mathfrak{f}_{\bf k}-P_{\rm free}.
\end{align}
Note, the term $-\Delta^2/4G_D$ (see Eq. (\ref{eq:Omega})) is neglected in this expression since it is small compared to the final result $P_{\rm pair}\propto\mu^2\Delta^2$.
Naive perturbative treatment of the pairing effects assumes expanding Eq. (\ref{eq:PCFLL1}) in powers of the pairing gap amplitude.
However, in the present case this procedure is ill defined since already in the leading order correction expansion coefficient diverges logarithmically due to the presence of the factor $-\mathfrak{e}_{\bf k}^{-1}$ under the momentum integral. 
This requires another expansion parameter, which is proportional to the pairing gap and provides convergence of the expansion coefficients. 
In order to define such parameter we notice that the main contribution to the momentum integral in Eq. (\ref{eq:PCFLL1}) is due to the momenta close by the absolute value to $\mu$.
In this case $\mathfrak{e}_{\bf k}$ is small and $\mathfrak{e}_{\bf k}\Delta_{j{\bf k}}$ can be used as a proper expansion parameter.
In order to make the next step we notice that in the leading order 
\begin{align}
\sqrt{\mathfrak{e}_{\bf k}^2+\Delta_{j{\bf k}}^2}=&
\sqrt{(\Delta_{j{\bf k}}-\mathfrak{e}_{\bf k})^2+2\mathfrak{e}_{\bf k}\Delta_{j{\bf k}}}
\nonumber\\
=&
\Delta_{j{\bf k}}-\mathfrak{e}_{\bf k}+
\mathcal{O}\left(\mathfrak{e}_{\bf k}\Delta_{j{\bf k}}\right).
\end{align}
It is important that the leading order correction in this expression includes the factor $\mathfrak{e}_{\bf k}$, which provides convergence of the momentum integrals at the upper limit of integration.
Performing some algebra with Eq. (\ref{eq:PCFLL1}) and using the above expansion we arrive at
\begin{align}
P_{\rm pair}=&
\sum_jd_j\int \frac{d{\bf k}}{(2\pi)^3}
\frac{\Delta_{j{\bf k}}^2\mathfrak{f}_{\bf k}}{\sqrt{\mathfrak{e}_{\bf k}^2+\Delta_{j{\bf k}}}-
\mathfrak{e}_{\bf k}}
\nonumber\\
\label{eq:PCFLL2}
=&\sum_jd_j\int \frac{d{\bf k}}{(2\pi)^3}
\frac{\Delta_{j{\bf k}}^2\mathfrak{f}_{\bf k}}{\Delta_{j{\bf k}}-2\mathfrak{e}_{\bf k}}
+\mathcal{O}(\Delta^3).
\end{align}
The leading order term in this expression is proportional to the squared pairing gap amplitude as is expected for the ABPR amplitude.
At the same time, the order $\mathcal{O}(\Delta^3)$ correction in Eq. (\ref{eq:PCFLL2}) is regular due to the presence of the factor $\mathfrak{e}_{\bf k}$ under the momentum integral.
As we already mentioned, the main contribution to this integral is due to $|{\bf k}|\simeq\mu$.
This allows us to replace the factor $\Delta_{j{\bf k}}^2$ in Eq. (\ref{eq:PCFLL2}) by its value at $|{\bf k}|=\mu$, i.e. by $\zeta_j^2\Delta_{\rm eff}^2$ since in the considered case $k_F=\mu$.
Due to the same reason we approximate the integration measure as
\begin{align}
\label{measure}
d{\bf k}\simeq d(\Delta_{\bf k}-2\mathfrak{e}_{\bf k})
\frac{4\pi|{\bf k}|^2}
{\frac{\partial}{\partial|{\bf k}|}(\Delta_{\bf k}-2\mathfrak{e}_{\bf k})}
\biggl|_{|{\bf k}|=\mu}.
\end{align}
The denominator in this expression is introduced in order to compensate the factor arising form the differential $d(\Delta_{\bf k}-2\mathfrak{e}_{\bf k})$. 
This denominator is $-2+\mathcal{O}(\Delta)$. 
With this Eq. (\ref{eq:PCFLL2}) gets
\begin{align}
\label{eq:PCFLL3}
P_{\rm pair}=
\sum_j\frac{d_j\zeta_j^2\mu^2\Delta_{\rm eff}^2}{(2\pi)^2}\int_0^\mu
\frac{d(\Delta_{j{\bf k}}-2\mathfrak{e}_{\bf k})}{\Delta_{j{\bf k}}-2\mathfrak{e}_{\bf k}}
+\mathcal{O}(\Delta^3).
\end{align}
The momentum integral in Eq. (\ref{eq:PCFLL3}) can be carried explicitly. It yields
$\ln(2\mu/\zeta_j\Delta_{\rm eff})+\mathcal{O}(1)$.
Thus, explicitly summing over the singlet and octet states we obtain
\begin{align}
\label{eq:PCFLL4}
P_{\rm pair}=\frac{3}{\pi^2}\mu^2\Delta_{\rm eff}^2
\left[\ln\frac{\mu^2}{\Delta_{\rm eff}^2}+\frac{\ln2}{6}\right]+
\mathcal{O}(\Delta^3).
\end{align}
Now we use the limit
$\lim_{x\to 0} x\,(\ln x^{-1}+c)= x$, which is derived in Appendix \ref{sec-app-a}
for any constant $c$, for the case $x=\Delta_{\rm eff}^2/\mu^2$.
It allows us to suppress the square bracket in the previous expression.
As it was argued in Sec. \ref{sec:nlNJL}, in the range of chemical potentials typical for quark matter in NSs the effective pairing gap $\Delta_{\rm eff}$ can be substituted by its average value $\Delta^*$. 
With this the correction caused by quark pairing becomes
\begin{align}
\label{eq:PCFLL5}
P_{\rm pair}=\frac{3}{\pi^2}\mu^2\Delta^{*2}.
\end{align}
Finally, combining Eqs. (\ref{eq:ABPR}), (\ref{eq:ABPR1}), (\ref{eq:PCFLL5}) we obtain an effective EOS of paired quark matter, which has the well known form of the ABPR model \cite{Alford:2004pf}.
However, in our formulation the key parameter of this model, i.e. the pairing gap $\Delta^*$, is directly derived from the microscopic nlNJL approach and can be straightforwardly connected to the parameters of its Lagrangian.

\section{Hybrid EOS and hybrid stars}

In order to obtain a hybrid EOS with a (hyper)nuclear hadronic phase at low densities and a transition to the CFLL quark matter phase at high densities, we  shall employ the two-phase approach.
This means that in this work we do not aim at a unified description of quark-hadron matter where the hadronic phase would emerge when starting from a microscopic approach we describe the hadrons as bound states of quarks in going beyond the mean field approximation. Here we restrict ourselves to the latter and chose an appropriate relativistic density functional model to describe the hadronic phase.
The transition between both phases is obtained by a Maxwell construction.


\subsection{Maxwell construction with hadronic EOS} 

We consider as hadronic EOS the relativistic density functional DD2npY-T with nucleons and hyperons \cite{Shahrbaf:2022upc}, which agrees with the low density constraint from the chiral effective field theory \cite{Kruger:2013kua}.

For the quark matter phase, we employ the ABPR EOS for which we fix the parameter $a_4=$\,\num{0.363} and vary both remaining free parameters, $\Delta^*$ and $B_{\rm eff}$.
Performing the Maxwell construction results in a set of hybrid EOS with a first order phase transition that occurs at a critical value of the baryonic chemical potential $\mu_c$ and pressure $P_c$ obtained from the Gibbs condition of pressure balance 
\begin{equation}
    \label{eq:Gibbs}
    P_c\equiv~P_h (\mu _c) = P_q (\mu _c)~.
\end{equation}
Note, for a given $\mu_c$ hadronic EOS unambiguously defines $P_c$ and critical energy density $\varepsilon_c\equiv\varepsilon_h(\mu_c)$.
Since nature prefers the EOS with the higher pressure,
the physical pressure at $\mu_B<\mu_c$ is that of the hadronic phase and at $\mu_B>\mu_c$, the hybrid EOS is given by the pressure of the ABPR quark matter model.

\begin{figure}[t]
    \centering
    \includegraphics[width=0.9\columnwidth]{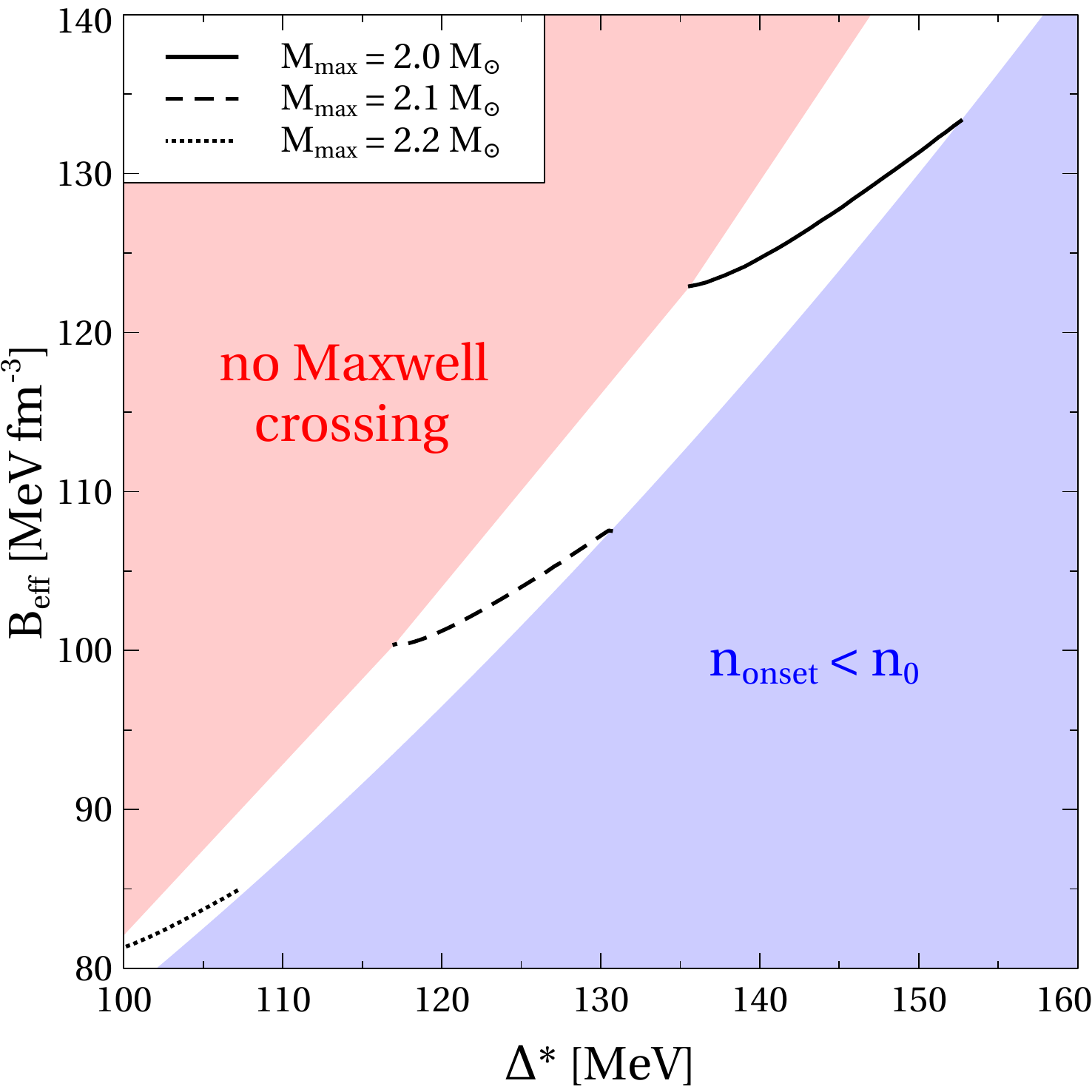}
    \caption{Admissible free parameters of the ABPR model for a hybrid EOS with DD2npY-T hadronic phase. Between inaccessible regions due to impossibility of a Maxwell construction (red area) and too low onset of deconfinement (blue area), the attainable maximum masses are shown by solid ($2.0\,{\rm M}_\odot$), dashed ($2.1\,{\rm M}_\odot$) and dotted ($2.2\,{\rm M}_\odot$) lines in the white area.}
    \label{fig:B-gap}
\end{figure}

In realizing the Maxwell construction, we observe that there is only a small corridor for the free parameters of the ABPR model, see Fig.~\ref{fig:B-gap}. 
The strong limitation stems from inaccessible regions in the 2D parameter space due to impossibility of a Maxwell construction (red area) and a too low onset of deconfinement (blue area). 
In between, there is a white area where hybrid EOS can be found that lead to reasonable hybrid NS sequences.

We note that replacing the Maxwell construction by a crossover interpolation scheme \cite{Kapusta:2021ney} removes the strong constraints on the parameter choice and allows to construct hybrid EOS with color superconductivity that fulfill the observational constraints in a wider  parameter range of the ABPR model
\cite{Blaschke:2021poc}.

\subsection{Calculation of astrophysical observables}

There is a one-to-one relationship between an EOS of dense matter and a sequence of NS configurations in the mass-radius diagram which is obtained by solving the Tolman-Oppenheimer-Volkoff (TOV) equations for the spherically symmetric (non-rotating) case \cite{PhysRev.55.364,PhysRev.55.374}.
These solutions can directly be compared to measurements of mass and radius, e.g., from the combined observations by NICER and XMM Newton of the millisecond pulsar J0740+6620 (see the analysis of Miller et al.\,\cite{Miller:2021qha}).
The mass of this object, $2.08\pm 0.07\,{\rm M}_\odot$ \cite{Fonseca:2021wxt}, sets a lower limit for the maximum mass that obtained for a given EOS by solving the TOV equations.
Additionally, the tidal deformability of a \num{1.4}\,${\rm M}_\odot$ NSs has been extracted from the gravitational wave measurement of the binary NS merger event GW170818. 
It has been obtained as $70<\Lambda<580$ \cite{LIGOScientific:2018cki}. 
The theoretical values of tidal deformability of NSs in dependence of their mass is obtained from solving a system of differential equations with the EOS as an input.

The astrophysical observables were calculated on the basis of the code by Andrea Maselli \cite{maselli2017}.

\begin{figure*}[!th]
\subfloat{
    \includegraphics[width=0.42\linewidth]{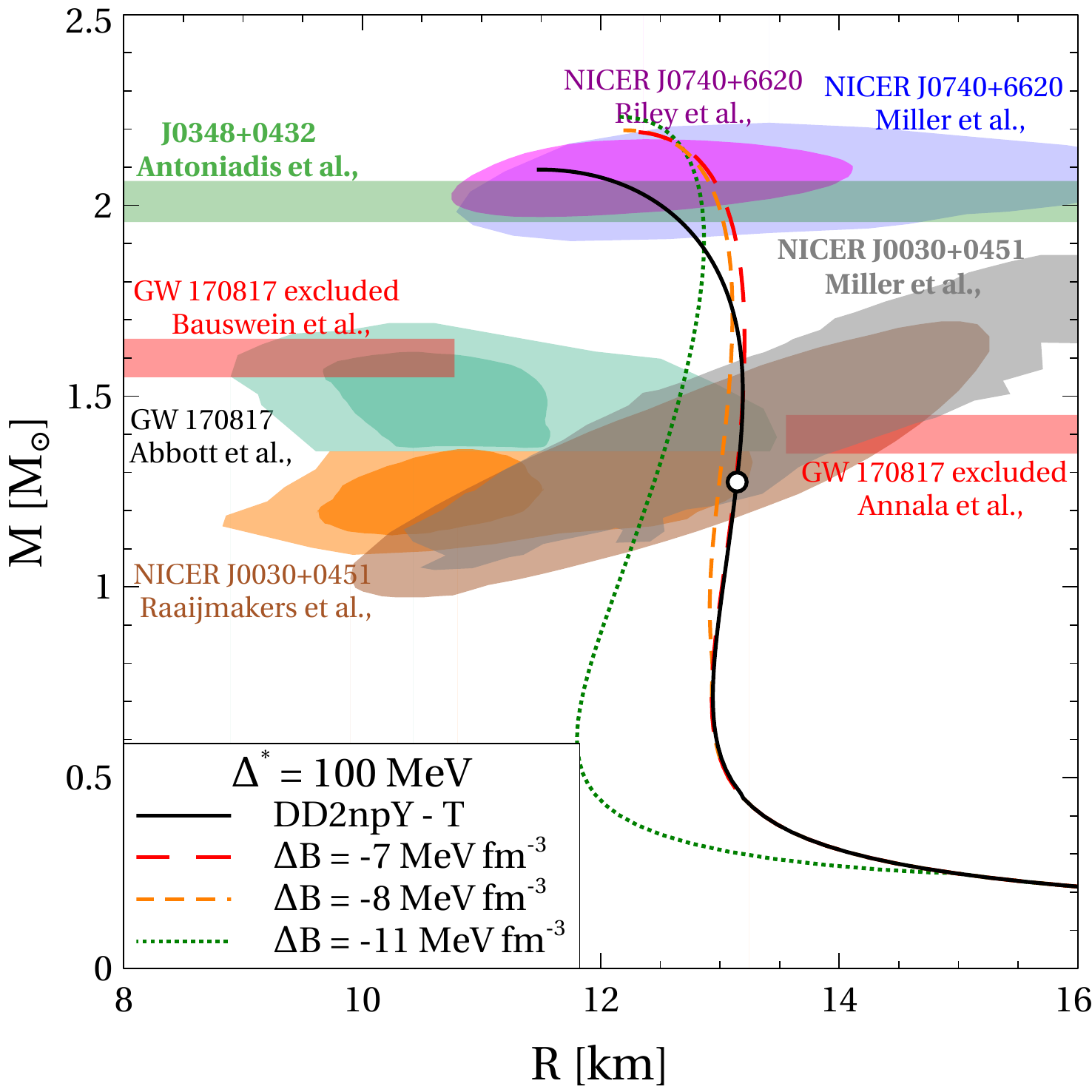} }\hspace{2em}
 \subfloat{
    \includegraphics[width=0.42\linewidth]{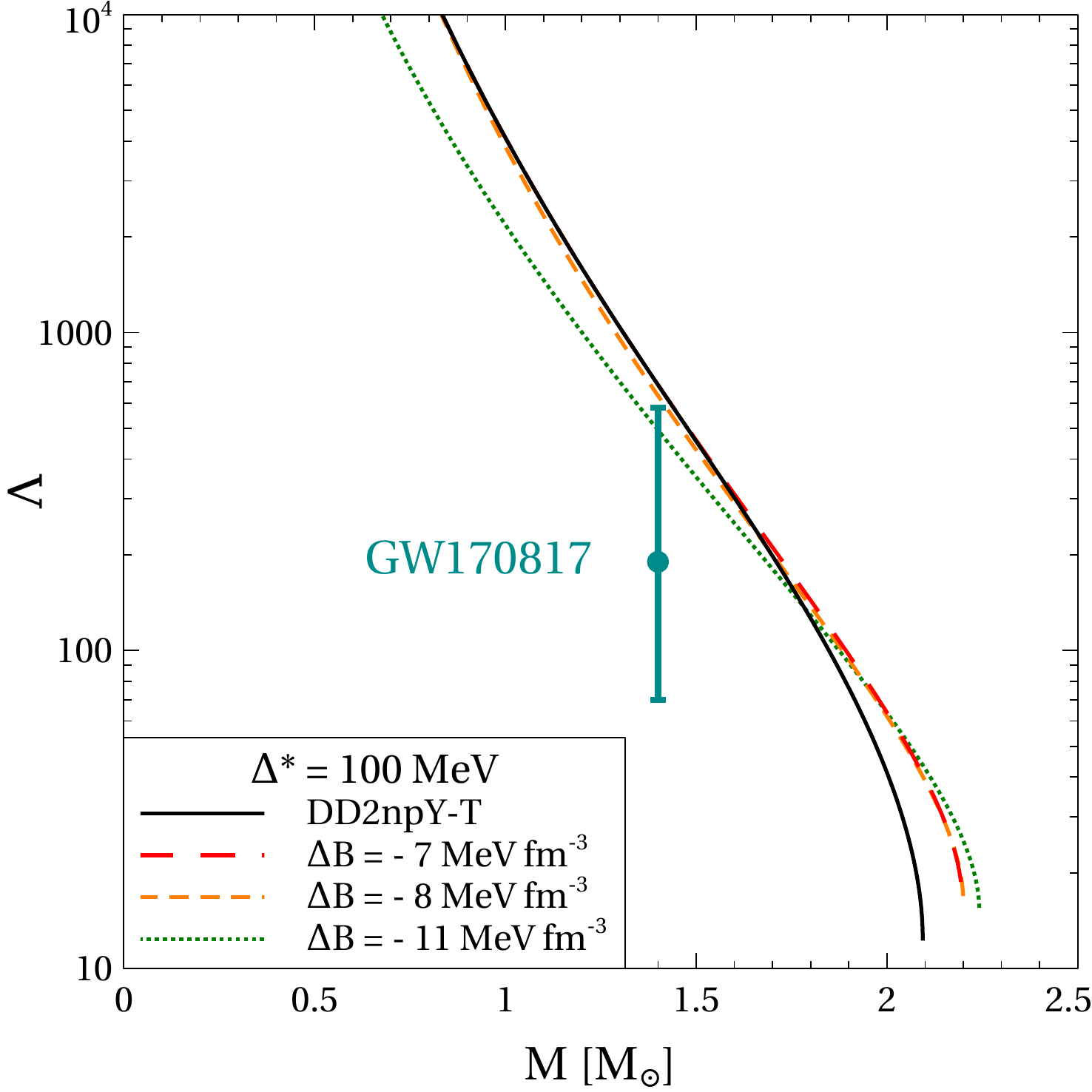} } \\
    \vspace{-1em}
\subfloat{
    \includegraphics[width=0.42\linewidth]{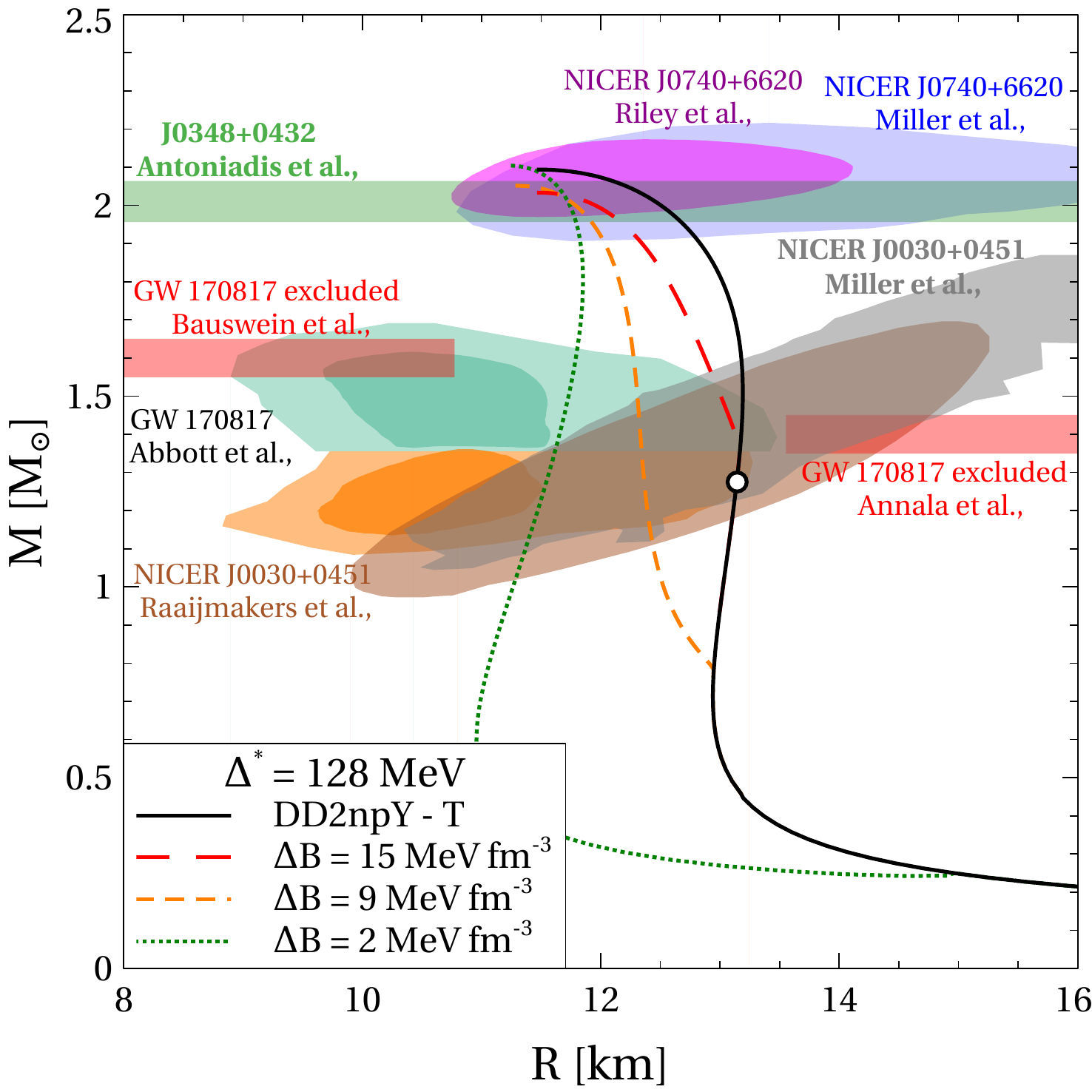} }\hspace{2em}
 \subfloat{
    \includegraphics[width=0.42\linewidth]{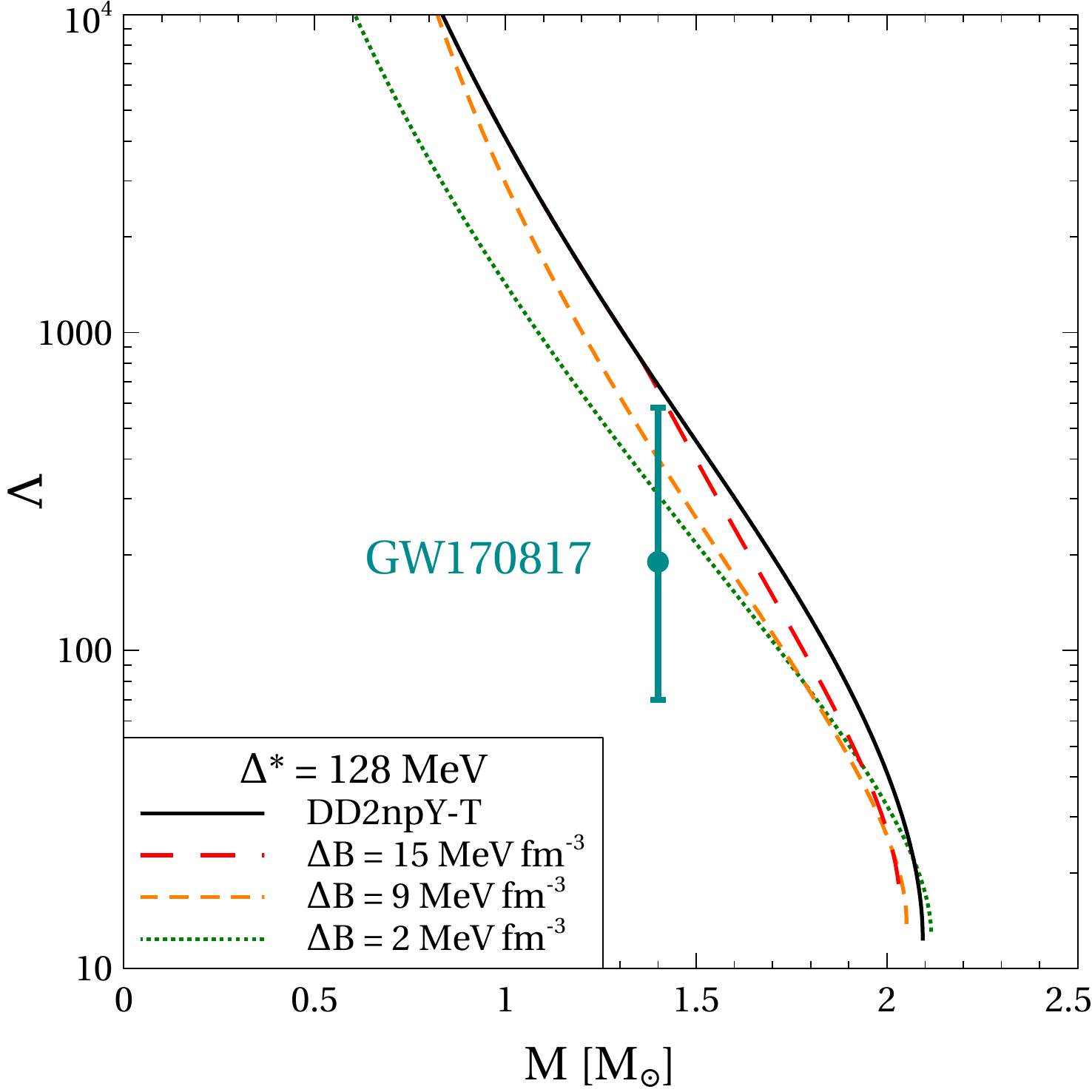} } \\
     \vspace{-1em}
\subfloat{
    \includegraphics[width=0.42\linewidth]{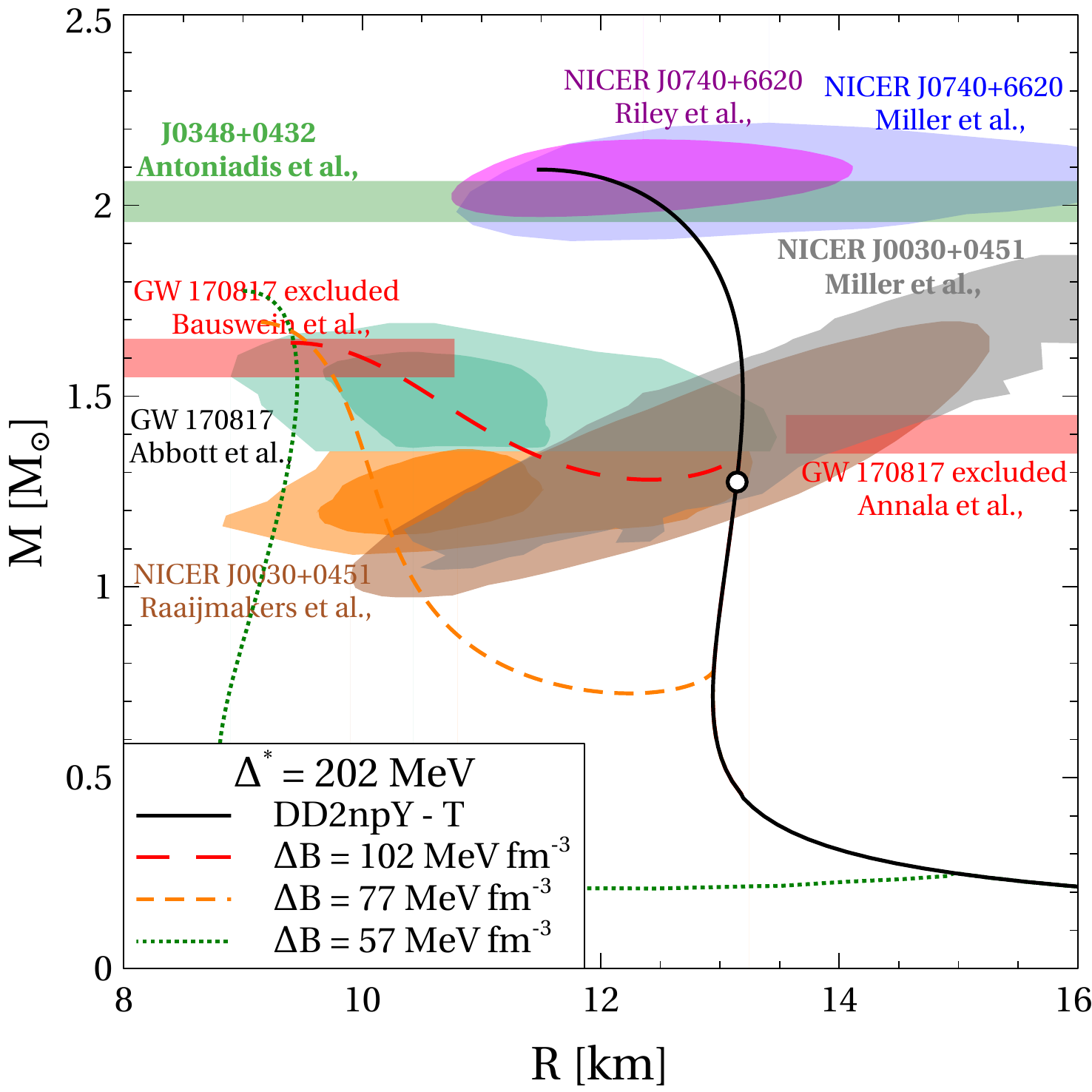}}\hspace{2em}
 \subfloat{
    \includegraphics[width=0.42\linewidth]{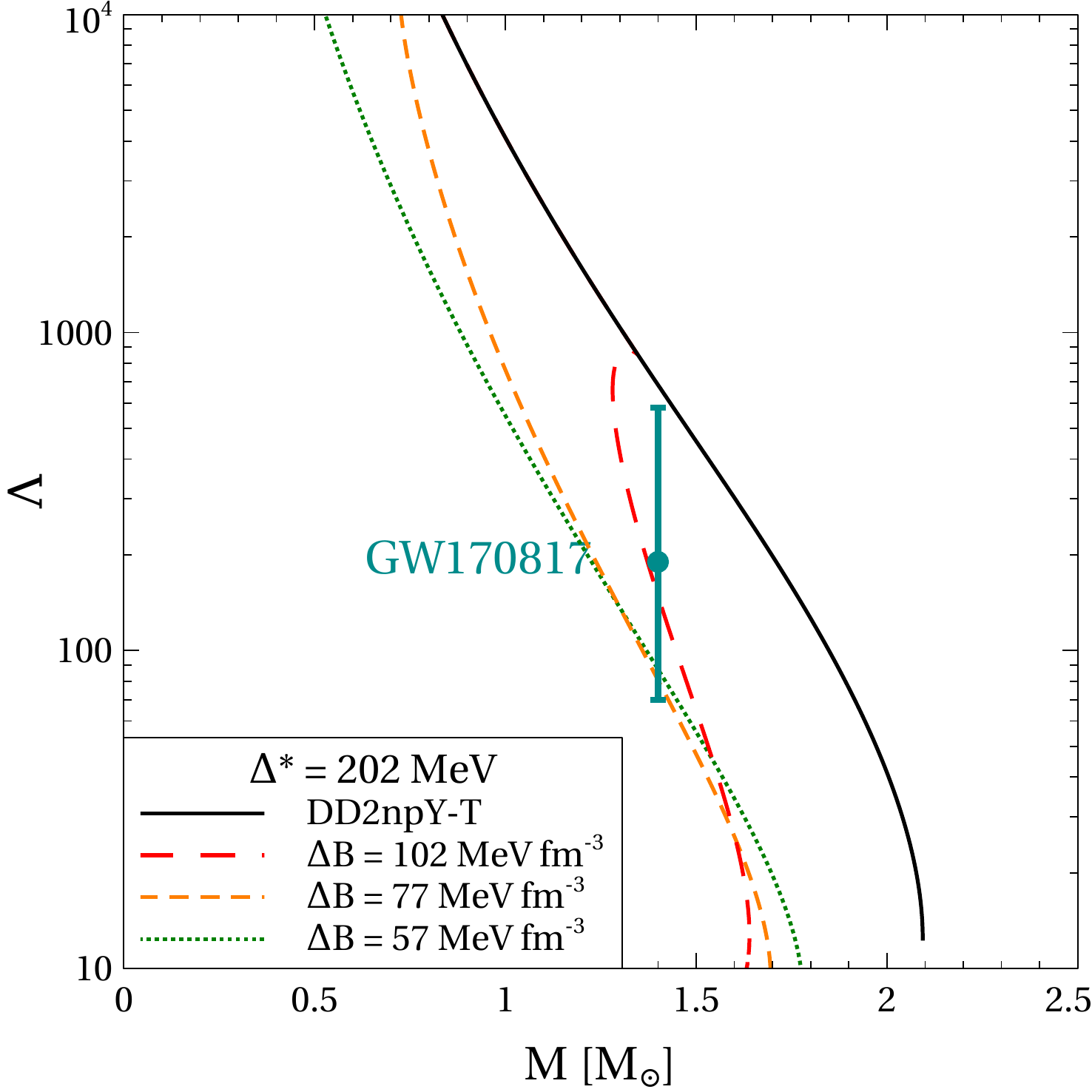}}
 \vspace{-1em}
     \caption{Left column: Mass-radius relation of hybrid NSs with the quark-hadron EOS obtained within the ABPR model with $\Delta^*=$\,\SI{100}{MeV} (upper panel), $\Delta^*=$\,\SI{128}{MeV} (middle panel) and $\Delta^*=$\,\SI{202}{MeV} (lower panel), corresponding to $\eta_D=$\,\num{0.45}, $\eta_D=$\,\num{0.5} and $\eta_D=$\,\num{0.6}, respectively. 
An empty circle on the hadronic curves indicates the hyperon onset.
The astrophysical constraints depicted by the colored bands and
shaded areas are discussed in the text.\\
Right column: 
Dimensionless tidal deformability $\Lambda$ as a function of the NS mass $\rm M$ (upper panel) obtained with the same hybrid EOS that were used in order to obtain the mass-radius relations shown in the left column. 
The observational constraint on the dimensionless tidal deformability of a $1.4\,{\rm M}_\odot$ NS is from the binary NS merger GW170817 
\cite{LIGOScientific:2018cki}.
}
    \label{fig:m-r}
\end{figure*}

\subsection{Observational constraints for EOS parameters}

The results for the sequences of star configurations in the mass-radius and tidal deformability-mass diagram
are shown in Fig.~\ref{fig:m-r} for three choices of the CFLL diquark pairing gap (upper, middle and lower panels) with the onset of deconfinement adjusted for each case to \numlist{0.24;0.84;1.37}\,${\rm M}_\odot$ by a proper choice of the corrective bag pressure parameter 
$\Delta B$. 
The characteristic values for mass, radius and tidal deformabilities are extracted from these solutions and given in table \ref{tab:param}.
The table also includes values of the baryonic chemical potential $\mu_B^{\rm max}$ and energy densities $\varepsilon^{\rm max}$ reached in the centers of the heaviest NSs with quark cores.
Larger diquark couplings or, equivalently, 
larger constant pairing gaps correspond to softer EOS of quark matter, which leads to larger values of $\mu_B^{\rm max}$ and $\varepsilon^{\rm max}$.
At the same time, for given values $\eta_D$ and $\Delta^*$ these central chemical potentials and energy densities depend only weakly on the onset mass $\rm M_{onset}$ for quark deconfinement, demonstrating a slightly rising behavior.

Table \ref{tab:param} indicates that an increase of $\Delta^*$ leads to a smaller maximum mass of the NS beccause of a softening of the ABPR EOS.
At first glance, this result contradicts the increase of the speed of sound for a larger pairing gap (see the lower panel of Fig. \ref{fig:cs2}) signalling the stiffening of the present EOS.
In order to resolve this apparent paradox, we should keep in mind that $\rm M_{max}$ is extracted form the solution of the TOV equation, which does not contain the speed of sound but pressure and energy density.
Therefore, within the TOV equation context it is reasonable to quantify the stiffness of an EOS not by the speed of sound $c_s^2=dp/d\varepsilon$, but by the dimensionless interaction measure $\delta=1/3-p/\varepsilon$.
Stiffer EOSs correspond to smaller $\delta$.
Inverting $c_s^2=c_s^2(\mu_B)$, the baryonic chemical potential can be eliminated from the expression for the interaction measure, which can be presented as a function of the speed of sound and bag pressure. 
A larger $c_s^2$ leads to smaller values of $\delta$, while increasing $B_{\rm eff}$ causes the opposite effect.
Using the Gibbs criterion (\ref{eq:Gibbs}) we express the latter as 
\begin{equation}
 \label{eq:Beff}
 B_{\rm eff}=\frac{3}{4\pi^2} a_4 \left(\frac{\mu_c}{3}\right)^4 + \frac{3}{\pi^2}\Delta^{*2} \left(\frac{\mu_c}{3}\right)^2 - P_c.
\end{equation}
This expression demonstrates that the effective bag pressure grows with $\Delta^*$, as the speed of sound does. 
As a result, the growth of the pairing gap induces two competing effects: an
decrease of $\delta$ caused by $c_S^2$ and its increase due to $B_{\rm eff}$.

\begin{figure}[t]
    \includegraphics[width=0.9\columnwidth]{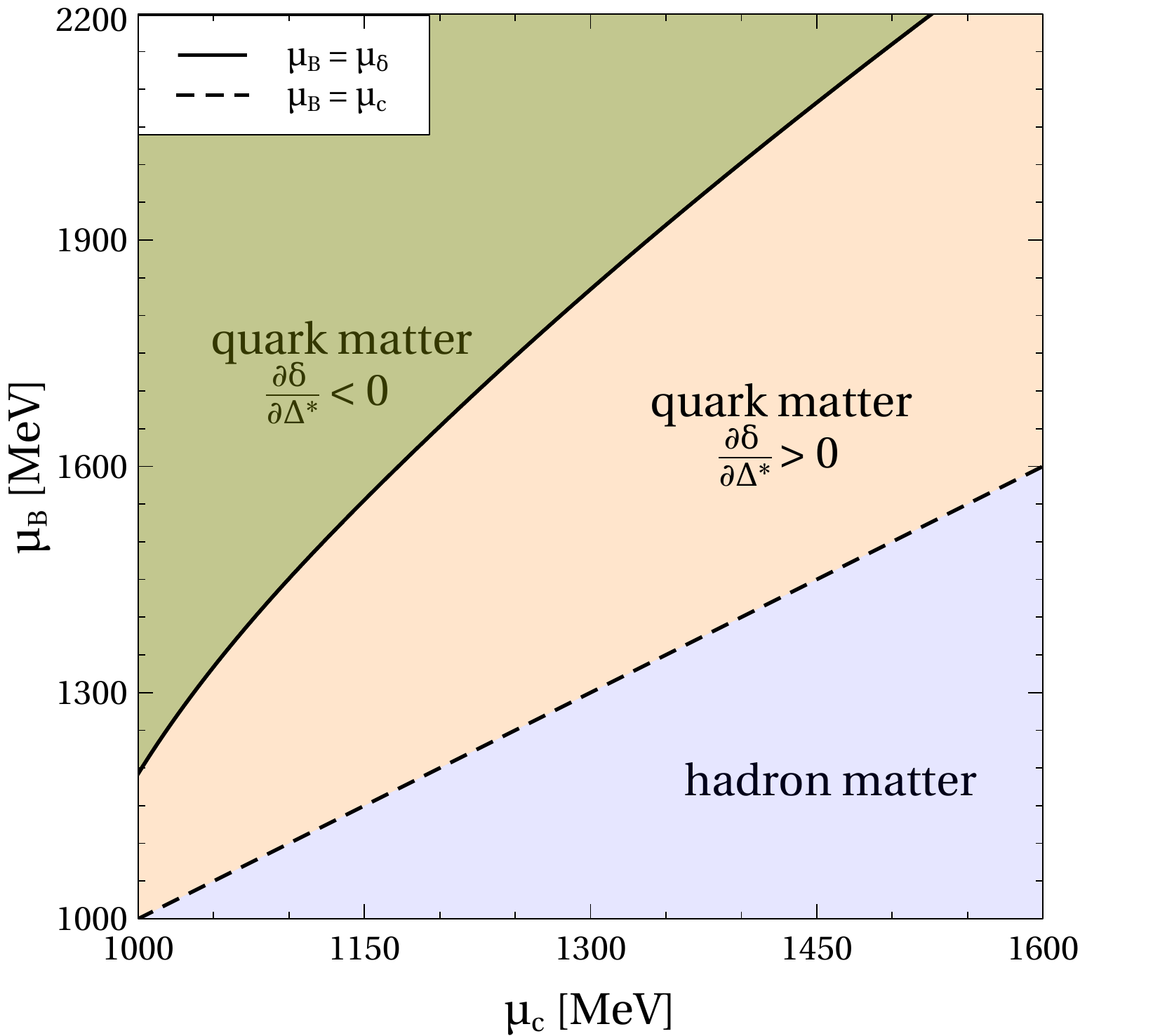}    \includegraphics[width=0.9\columnwidth]{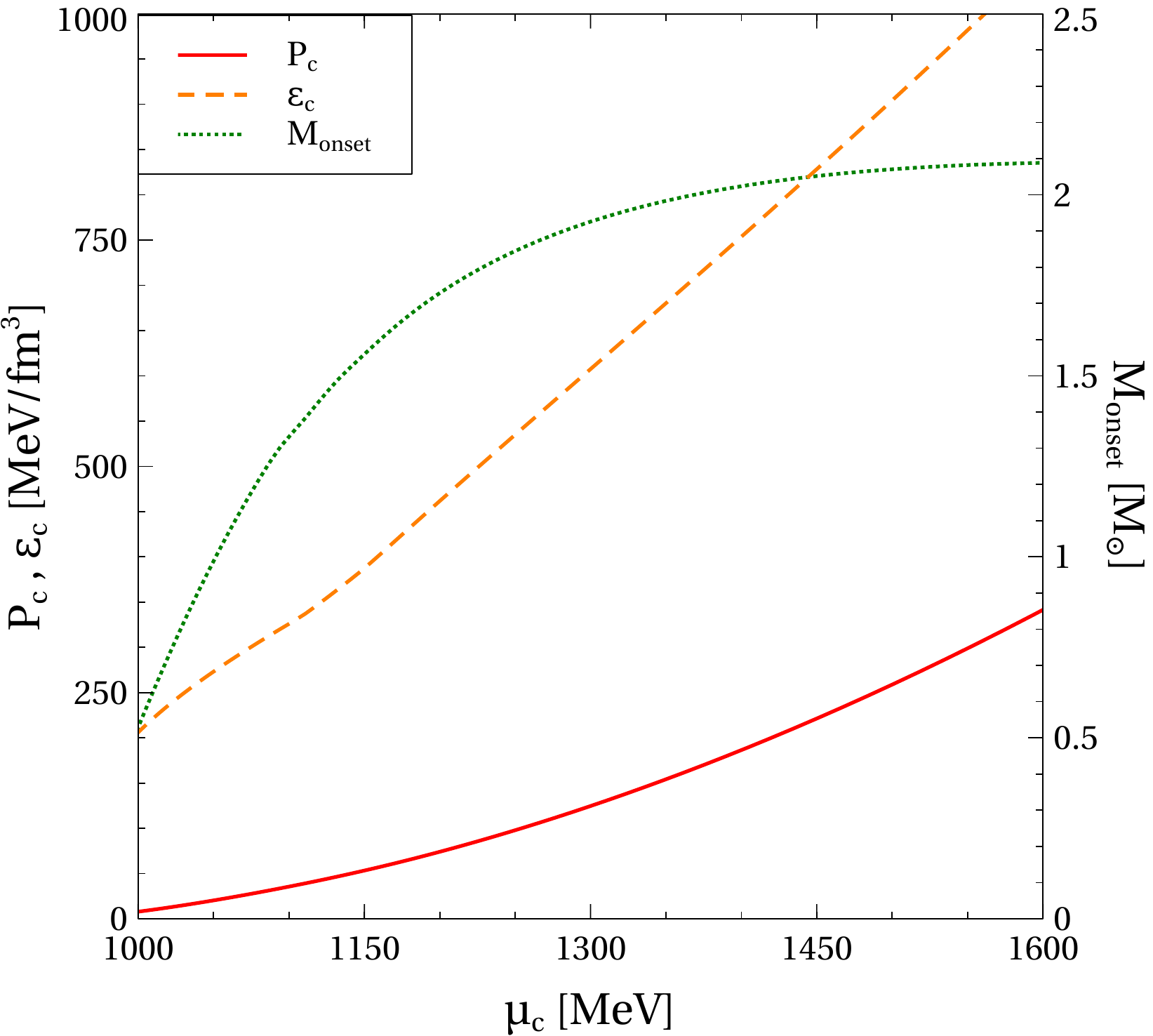}
    \caption{Ranges of the pairing gap induced softening ($\partial\delta/\partial\Delta^*>0$) and stiffening ($\partial\delta/\partial\Delta^*<0$) of the ABPR EOS, separated by $\mu_\delta$ (upper panel) as well as critical pressure $P_c$, energy density $\varepsilon_c$ and onset mass $\rm M_{onset}$ for deconfinement (lower panel) as a function of the critical chemical potential $\mu_c$.
    }
    \label{fig:mu_delta}
\end{figure}

In order to show that within the range of baryonic chemical potentials typical for NSs the second of these effects dominates, we consider the derivative
\begin{eqnarray}
\label{eq:derivative1}
\frac{\partial\delta~}{\partial\Delta^*}=
\frac{p}{\varepsilon^2}\frac{\partial\varepsilon}{\partial\Delta^*}-
\frac{1}{\varepsilon}\frac{\partial p}{\partial\Delta^*}.
\end{eqnarray}
Using explicit expressions for $p$, $\varepsilon$, $B_{\rm eff}$ and performing straightforward manipulations, Eq. (\ref{eq:derivative1}) becomes
\begin{eqnarray}
\label{eq:derivative2}
\frac{\partial\delta~}{\partial\Delta^*}&=&
\frac{12\Delta^*\mu^2}{\pi^2\varepsilon^2}
\left[P_c-\frac{3a_4}{4\pi^2}\left(\mu^2-\left(\frac{\mu_c}{3}\right)^2\right)^2\right].
\end{eqnarray}
From this expression we conclude that $\delta$ increases with $\Delta^*$ if the baryonic chemical potential is below 
\begin{eqnarray}
 \mu_\delta=\mu_c\sqrt{1+6\pi\sqrt{\frac{3P_c}{a_4\mu_c^4}}},
\end{eqnarray}
while at $\mu_B>\mu_\delta$ the interaction measure decreases with the pairing gap growth.
This corresponds to the pairing gap induced softening and stiffening of the present EOS, respectively.
Since $\mu_\delta$ always exceeds $\mu_c$, then $\delta>\delta|_{\Delta^*=0}$ after the hadron-to-quark matter phase transition, which corresponds to softening of the ABPR EOS. The corresponding ranges of the baryonic chemical potential are shown on the upper panel of Fig. \ref{fig:mu_delta}.
On the lower panel of Fig. \ref{fig:mu_delta}, we show the quantities which characterize the onset of deconfinement, the critical pressure $P_c$, the critical hadronic energy density $\varepsilon_c$ and the NS mass 
$M_{\rm onset}$ as a function of $\mu_c$. 

As a result of the present work and the dictionary we provided, one can constrain the values of the free parameters in the nlNJL model Lagrangian for dense QCD matter with the help of NS phenomenology.
To this end, we also indicated in Fig. \ref{fig:B-gap} 
the isolines of the achievable maximum mass for an admissible pair of ABPR EOS parameters $\Delta^*$ and $B_{\rm eff}$ (white region).
The larger the chosen $\Delta^*$, the softer the EOS becomes and the smaller is the achievable maximum mass. For $\Delta^*\,\gtrsim$\,\SI{150}{MeV}, the lower limit of the maximum mass cannot be reached. This is also illustrated in Fig. \ref{fig:m-r} for the case 
$\Delta^*\,=$\,\SI{202}{MeV}.
 From Fig. \ref{fig:Delta} one reads off that the upper limit for $\Delta^* \lesssim$\,\SI{150}{MeV} constrains the dimensionless diquark coupling to
$\eta_D \lesssim$\,\num{0.53}.
At the same time, for the dimensionless vector coupling holds $\eta_V\gtrsim$\,\num{1.2}.


\section{Conclusions}

In this work, we have derived the effective cold quark matter model by Alford, Braby, Paris and Reddy (ABPR) from a nonlocal Nambu--Jona-Lasinio model for the color superconducting quark matter in the CFLL phase with three light flavors.

We have discussed a generalization of the ABPR model that uses an $\mathcal{O}(\alpha_s)$ perturbative QCD correction with a running strong coupling constant $\alpha_s(\mu)$ that assures reaching the conformal limit for the squared speed of sound $c_s^2 - 1/3 \to 0^-$ for high chemical potentials. 
Below a matching point that is shown to lie above the range of chemical potentials that can be accessed in NSs, the running coupling is switched to a constant value $\alpha_s=$\,\num{1} which is large enough to provide the necessary stiffness of the quark matter phase for reaching maximum masses of hybrid stars in accordance with the observational lower limit ${\rm M}_{\rm max} \ge 2.01\,{\rm M}_\odot$. 

We have shown that due to the momentum dependence of the pairing which is induced by the nonlocality of the interaction, the effective gap parameter in the EOS model has a plateau-like behavior in the range of chemical potentials for NSs that is well approximated by a constant value. 
The dependence of this value on the diquark coupling strength in the nlNJL model Lagrangian could be fitted to a parabola.
NS phenomenology constrains this pairing gap parameter to values between \num{100} and \SI{150}{MeV} which translate to the narrow range of diquark couplings $\eta_D=0.45 \dots 0.53$. 
Due to this large pairing gap, the sound speed of the ABPR EOS exceeds the conformal limit value.
   
The dictionary for translating the vector meson and diquark coupling as free parameters of the nlNJL model to those of the ABPR model that is completed by relating the effective bag pressure parameter to the vector meson coupling. 
In order to fulfill also the low tidal deformability constraint from GW170817, a softening of the EOS on the hybrid NS branch is necessary which requires an early onset of quark deconfinement at 
${\rm M}_{\rm onset}<1.4~{\rm M}_\odot$.
To assure the early onset, a small pressure correction of $\Delta B \sim$\,\SI{10}{\mega eV\per\femto\meter\tothe3} is required which could be justified by a modification of the nonperturbative gluon sector at high baryon densities.

Summarizing, we have provided a microphysical justification for the use of the ABPR EOS in NS phenomenology based on the nonlocal NJL model for color superconducting quark matter in the CFLL phase.
A dictionary is provided for relating the 
ABPR EOS parameters for which constraints from the analysis of NS phenomenology are fulfilled to the free parameters of the nlNJL model Lagrangian for low-energy QCD.
We find that a finite pairing gap corresponds to a squared sound speed that exceeds the conformal limit. Increasing the diquark coupling and thus the pairing gap, however, softens the EOS and entails a lowering of the maximum mass. An optimal diquark pairing gap for which the maximum mass exceeds $2~M_\odot$ is of the order of 100 \dots 120  MeV.

There are several routes one could follow in subsequent work, based on the present study.
In concluding, we would like to mention a few of them.
The assumption of the degeneracy of the strange quark mass with that of the up and down quarks could be relaxed. 
Then, one could follow the route of the nonlocal NJL model for that case, or one could employ the approach of the confining density functional \cite{Ivanytskyi:2022oxv} which has been recently developed by two of us (D.B., O.I.). 
The main difference would be in the phase structure that is to be expected, namely the existence or non-existence of a two-flavor color superconducting (2SC) phase and the possibility to address in a microscopic model the case of absolutely stable strange quark matter.
Besides the simple Maxwell construction of a hadron-to-quark matter transition, a crossover transition could be constructed, eventually with a corridor of a first-order transition and two critical endpoints.
The appropriate method of a two-zone interpolation scheme \cite{Ivanytskyi:2022wln}
has been developed by two of us (O.I., D.B.).
Also, the simple ansatz of the switch model for the running coupling could be systematically developed. As a first step, one could choose different values of the saturated (constant) coupling in the nonperturbative domain. This would change the switch point and eventually lead to an intrusion of the region where the coupling is running to the NS density domain. Such a variation of the $a_4$ parameter of the ABPR model would then be reflected in a variation of the $\eta_V$ parameter of the nlNJL model independently of the $\eta_D$ parameter.
A more realistic ansatz for the running of the QCD coupling in the nonperturbative domain \cite{Deur:2016tte} could be chosen. Then, a direct influence of the detailed scheme of such a running coupling like, e.g., that of the analytic perturbation theory, on the NS phenomenology would be expected.

\subsection*{Acknowledgements}
The work of D.B., O.I. and U.S. has been supported in part by the Polish National Science Centre (NCN) under grant No. 2019/33/B/ST9/03059.
The authors acknowledge the COST Action CA16214 “PHAROS” for the supporting their networking activities, in particular the PHAROS Training School on "Equation of State of Dense Matter and Multi-Messenger Astronomy" in Karpacz, June 13-19, 2021.

\begin{appendix}
\section{Limit of Eq. (30)}
\label{sec-app-a}
Here for the readers convenience we derive the relation 
\begin{align}
\label{eq:A1}
\lim_{x\rightarrow0} x^a\left(\ln\frac{1}{x} +c\right)=\lim_{x\rightarrow0} 
\frac{x^a}{a}.
\end{align}
In Sec. \ref{ssec:matching} it is used at $a=1$, while for the sake of generality we derive it for any nonvanishing $a$. 
At the first step we rewrite $x^a=1/x^{-a}$.
This yields an indeterminate form $(\frac{\infty}{\infty})$.
It can be treated using the L'H\^opital's rule assuming replacement of $\ln\frac{1}{x} +c$ and $x^{-a}$ by their derivatives.
Thus
\begin{align}
\label{eq:A2}
\lim_{x\rightarrow0} x^a\left(\ln\frac{1}{x} +c\right)=
\lim_{x\rightarrow0} \left(-\frac{1}{x}\right) \biggl/
\left(-\frac{a}{x^{a+1}}\right).
\end{align}
After a simple algebra this relation arrives at the desired form of Eq. (\ref{eq:A1}). 
\label{sec-app-a}

\end{appendix}
\bibliography{qcp}

\end{document}